\documentclass[12pt]{article}
\pdfoutput=1
\usepackage{amsthm, amsmath,amsfonts,graphicx,color,bbm,tikz}
\usepackage{comment}
\usepackage[nosort]{cite}
\usepackage{subfigure}
\usetikzlibrary{calc,positioning}
\usetikzlibrary{patterns,arrows,decorations.pathreplacing}
\usepackage{caption}
\usepackage{ulem}
\tikzset{>=stealth}
\usepackage{lipsum}
\usepackage{tabularx}
\usepackage{pgfplots}
\pgfplotsset{compat=1.13}
\usepackage{braids}
\usetikzlibrary{braids}						% TeX will automatically convert eps --> pdf in pdflatex	
\usepackage{fullpage}
\usepackage{hyperref}
\usepackage{tikz}
\usetikzlibrary{quantikz}
\newtheorem{thm}{Theorem}

\newtheorem{definition}{Definition}

\makeatletter\@addtoreset{equation}{section}\makeatother

\newcommand{\be}{\begin{equation}}
\newcommand{\ee}{\end{equation}}
\def\beq{\begin{equation}}
\def\eeq{\end{equation}}
\newcommand{\bea}{\begin{eqnarray}}
\newcommand{\eea}{\end{eqnarray}}
\newcommand{\Tr}{{\rm Tr\,}}

\renewcommand{\title}[1]{\vbox{\center\LARGE{#1}}\vspace{3mm}}
\renewcommand{\author}[1]{\vbox{\center{#1}}\vspace{3mm}}

\newcommand{\email}[1]{\vbox{\center\tt#1}\vspace{3mm}}

\begin{document}
\begin{titlepage}

\begin{center}
{\large {\bf Dyck Paths and Topological Quantum Computation } }

\author{ Vivek Kumar Singh, Akash Sinha,
Pramod Padmanabhan, Indrajit Jana  }

%\vskip 0.25cm
{{\it Center for Quantum and Topological Systems (CQTS), NYUAD Research Institute, \\ New
York University Abu Dhabi, PO Box 129188, Abu Dhabi, UAE}}\\
{{\it School of Basic Sciences,\\ Indian Institute of Technology, Bhubaneswar, 752050, India}}

\email{vks2024@nyu.edu, akash.sinha, pramod23phys, jana.indrajit@gmail.com }

\vskip 3cm 

\end{center}

%%%%%%%%%%%

\abstract{
\noindent 
The fusion basis of Fibonacci anyons supports unitary braid representations that can be utilized for universal quantum computation. 
%The 2-strand and 3-strand braid groups, $\mathcal{B}_2$ and $\mathcal{B}_3$, obtained this way are irreducible, unitary and are used to construct the higher strand braid groups on this space. 
We show a mapping between the fusion basis of three Fibonacci anyons, $\{\ket{1}, \ket{\tau}\}$, and the two length 4 Dyck paths  {\it via} an isomorphism between the two dimensional braid group representations on the fusion basis and the braid group representation built on the standard $(2,2)$ Young diagrams using the Jones construction. This correspondence helps us construct the fusion basis of the Fibonacci anyons using Dyck paths as the number of standard $(N,N)$ Young tableaux is the Catalan number, $C_N$ . We then use the local Fredkin moves to construct a spin chain that contains precisely those Dyck paths that correspond to the Fibonacci fusion basis, as a degenerate set. We show that the system is gapped and examine its stability to random noise thereby establishing its usefulness as a platform for topological quantum computation. Finally, we show braidwords in this rotated space that efficiently enable the execution of any desired single-qubit operation, achieving the desired level of precision($\sim 10^{-3}$).

}

\end{titlepage}
\tableofcontents 

%%%%%%%%%%%%%%%%%%%%%%%%%%%%%%%%%%%%%%%%%%%%%%%%%%%%%%%%%%%%%%%%%%%%%%%%%%%%%%
\section{Introduction}
\label{sec:Introduction}
%%%%%%%%%%%%%%%%%%%%%%%%%%%%%%%%%%%%%%%%%%%%%%%%%%%%%%%%%%%%%%%%%%%%%%%%%%%%%%
Topology and its accompanying effects provide a promising route to overcome the fragile nature of quantum states. This is best seen in {\it topological quantum computation} \cite{wang2010topological,freedman2003topological,Lahtinen_2017,nayak2008non,pachos2012introduction}, realised by building quantum gates using the elements of the braid group. The central players in this theory are non-Abelian {\it anyons}, a class of quasiparticles found in two dimensional systems \cite{Bonderson2007NonAbelianAA}. In more technical terms these anyons are the representations of the braid group $\mathcal{B}_N$ which are the superselected groups in two dimensional systems as opposed to the permutation group in higher dimensions \cite{fredenhagen1989superselection,rehren1989braid}. While there are many proposals for obtaining non-Abelian anyons in physical systems \cite{lesanovsky2012interacting,stoudenmire2015assembling,vaezi2014fibonacci,mong2017fibonacci,djuric2017fibonacci} they have so far been elusive in experiments. In this scenario it is important to find alternate sources that either produce such exotic particles or mimic their effects. In the latter vein we recently proposed supersymmetric (SUSY) spin chain systems \cite{Jana2022TopologicalQC} on which the braid group of certain non-Abelian anyons was realised {\it via} the Jones representation \cite{c0350d72-fa69-31c2-9f82-f38a198cf1f6}. In particular we found SUSY systems that supported the braid groups corresponding to the Ising, Fibonacci and Jones-Kauffman anyons \cite{kauffmanlomonaco,https://doi.org/10.48550/arxiv.1501.02841,PhysRevA.92.012301} which form the IRR's of the $SU(2)_k$ quantum groups for $k=2,3,4$ \cite{Biedenharn:1996vv, Majid1995FoundationsOQ} respectively. In this paper we carry this idea forward by generalizing it to a spin chain built out of the {\it Fredkin moves} \cite{Fredkin2002ConservativeL}. 
The Fredkin moves are local operators that connect {\it Dyck paths} which are one dimensional paths that begin and end on the line but never go below the line. The path is composed of `up' and `down' steps which can be interpreted as a two-level or qubit system on each step. Spin chains made out of these moves are known to exhibit interesting properties ranging from entanglement entropies which violate the area law, phase transitions and localisation \cite{Salberger2016FredkinSC,Padmanabhan2018QuantumPT,Zhang2017EntropyGA,Salberger2016DeformedFS,Langlett2021HilbertSF,Zhang2016NovelQP}. Here we construct yet another spin chain using these moves that can support the Fibonacci braid group. The premise for this construction is based on the identification of a unitary transformation between two different representations of the braid group. In this work we explain this and study its consequences. 

We begin with an observation that the two dimensional representation of $\mathcal{B}_3$ on the Fibonacci anyon fusion basis is isomorphic to the $(2,2)$ Young tableaux representation of $\mathcal{B}_4$\footnote{In the $(2,2)$ Young tableaux representation of $\mathcal{B}_4$, $\sigma_1=\sigma_3$ and hence the isomorphism makes sense.}, obtained {\it via} the Jones representation. Using this we establish a correspondence between the Fibonacci anyon fusion basis, $\{\ket{1}, \ket{\tau}\}$ and the standard $(2,2)$ Young diagrams. Then we note that the number of standard $(N,N)$ Young diagrams equals the {\it Catalan number}, $C_N$ and thus we can establish a one-to-one mapping between these Young diagrams and the length $2N$ Dyck paths. First we identify the fusion basis of three Fibonacci anyons with the two length four Dyck paths. This is then extended to a chain of arbitrary length where the fusion basis is mapped to a restricted set of height 2 Dyck paths. Following this we construct the spin chain which projects out precisely these states as a degenerate set. The projectors are local and are constructed using the Fredkin moves. We then show that this system has an energy gap which is stable to random noise using known theorems from random matrix theory. Finally we analyse the braid words in this setup that approximate a universal set of gates required for quantum computing. Universality is expected as we have a unitary transformation of the Fibonacci braid group. Our findings lead us to the conclusion that it is possible to efficiently identify braids that can execute any desired single-qubit operation with the desired level of accuracy. Additionally, we provide evidence \cite{Bernard-Simula} that the particular braid words associated with target gates have shorter lengths when compared to the Fibonacci case, even within the same error ranges(equivalent to $\sim$ $10^{-3}$).

These ideas are arranged as follows. We start with a review of the representation theory of the {\it Temperley-Lieb algebra} \cite{Temperley1971RelationsBT} and the braid group built using this in Sec. \ref{sec:isomorphism}. This section also includes the two dimensional IRR's that are shown to be unitarily equivalent to the appropriate braid groups built on the Fibonacci fusion spaces. The spin chain that project out this fusion basis is constructed using the local Fredkin moves in Sec. \ref{sec:spinchain} and its gap and stability are analysed in Secs. \ref{sec:energygap} and \ref{sec:stability} respectively. The universal computational properties of this system is studied in Sec. \ref{sec:TQC}. We end with a scope of this construction in Sec. \ref{sec:outlook}.

%%%%%%%%%%%%%%%%%%%%%%%%%%%%%%%%%%%%%%%%%%%%%%%%%%%%%%%%%%%%%%%%%%%%%%%%%%%%%%
\section{The Isomorphism}
\label{sec:isomorphism}
%%%%%%%%%%%%%%%%%%%%%%%%%%%%%%%%%%%%%%%%%%%%%%%%%%%%%%%%%%%%%%%%%%%%%%%%%%%%%%
The motivation for our construction originates from an isomorphism between the Jones form of the braid group representation on two different spaces. The first space consists of standard Young diagrams of the $(2,2)$ type and the second space is spanned by the fusion basis elements that appear when we fuse three Fibonacci anyons. Following this we show that there is a one-to-one correspondence between the standard Young diagrams of the $(N,N)$ type and the length $2N$ Dyck paths leading us to the desired mapping between the length 4 Dyck paths the fusion basis of three Fibonacci anyons. The three ingredients required for establishing this mapping are known results which we will first recall.

%%%%%%%%%%%%%%%%%%%%%%%%%%%%%%%%%%%%%%%%%%%%%%%%%%%%%
\subsection*{Jones representation of $\mathcal{B}_N$}
\label{subsec:Jonesrep}
\begin{figure}[h]
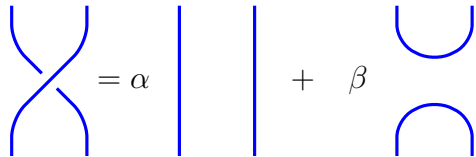

	\centering
	\begin{align*}
	\tikz[baseline=-6ex, scale=1]{
		\def\a{0}
		\def\b{0}
		\draw[very thick, blue, rounded corners=7] (\a + 1, \b) -- (\a + 1, \b - 0.5) -- (\a + 1.4, \b - 0.9);
		\draw[very thick, blue, rounded corners=7] (\a + 1.6, \b - 1.1) -- (\a + 2, \b - 1.5) -- (\a + 2, \b - 2);
		\draw[very thick, blue, rounded corners=7] (\a + 2, \b) -- (\a + 2, \b - 0.5) -- (\a + 1, \b - 1.5) -- (\a + 1, \b - 2);
	} = \alpha
	\tikz[baseline=-6ex, scale=1]{
		\def\a{0}
		\def\b{0}
		\node at (\a - 0.25, 0) {};
		\draw[very thick, blue, rounded corners=7] (\a, \b) -- (\a, \b - 2);
		\draw[very thick, blue, rounded corners=7] (\a + 1, \b) -- (\a + 1, \b - 2);
		\node at (\a + 1.25, 0) {};
	}
	+\;\;\;
	\beta
	\tikz[baseline=-6ex, scale=1]{
		\def\a{0}
		\def\b{0}
		\node at (\a - 0.25, 0) {};
		\draw[very thick, blue, rounded corners=7] (\a, \b) -- (\a, \b - 0.5) -- (\a + 0.5, \b - 0.75) -- (\a + 1, \b - 0.5) -- (\a + 1, \b);
		\draw[very thick, blue, rounded corners=7] (\a, \b - 2) -- (\a, \b - 1.5) -- (\a + 0.5, \b - 1.25) -- (\a + 1, \b - 1.5)-- (\a + 1, \b - 2);
	}
	\end{align*}
	\caption{Jones representation of the braid group in pictures.}
	\label{fig:Jonesrep}
\end{figure}

%%%%%%%%%%%%%%%%%%%%%%%%%%%%%%%%%%%%%%%%%%%%%%%%%%%%%
We can build the generators of the braid group $\mathcal{B}_N$ using the Temperley-Lieb algebra, $TL_N(x)$ as, 
\begin{equation}\label{eq:sigmaJonesrep}
    \sigma_i = \alpha~1 + \beta~e_i,
\end{equation}
where $\alpha$ and $\beta$ are complex numbers satisfying, $\alpha^2 + \beta^2 + x\alpha\beta=0$. The operators, $e_i$ are the generators of $TL_N(x)$ and they obey the relations,
\begin{equation}\label{eq:TLrelations}
    e_i^2 = x~e_i,~~e_ie_{i+1}e_i = e_{i+1}e_ie_{i+1},~~e_ie_j=e_je_i,~\textrm{when}~|i-j|>1.
\end{equation}
Given these it is simple to verify the braid generators in \eqref{eq:sigmaJonesrep} follow,
\begin{equation}\label{Eq:braidrelations}
    \sigma_i\sigma_{i+1}\sigma_i = \sigma_{i+1}\sigma_i\sigma_{i+1},~~\sigma_i\sigma_j = \sigma_j\sigma_i,~\textrm{when}~|i-j|>1,
\end{equation}
thus providing a representation of the $N$-strand braid group, $\mathcal{B}_N$. This is known as the Jones representation \cite{c0350d72-fa69-31c2-9f82-f38a198cf1f6} and in general they are not unitary and may not realize the full infinite dimensional braid group. The pictorial form of this representation (See Fig. \ref{fig:Jonesrep}) is instrumental in the state-sum construction of the Kauffman bracket which are knot and link invariants \cite{kauffman1987state}.
%%%%%%%%%%%%%%%%%%%%%%%%%%%%%%%%%%%%%%%%%%%%%%%%%%%%

As is evident from this definition, the Temperley-Lieb algebra plays an important role in its construction. So we will now show two ways of constructing the representations of the Temperley-Lieb algebras useful for our purposes.

%%%%%%%%%%%%%%%%%%%%%%%%%%%%%%%%%%%%%%%%%%%
\subsection*{Young Tableaux representation}
\label{subsec:YTrep}
\begin{figure}[h]
     \centering
     \includegraphics[width=8cm]{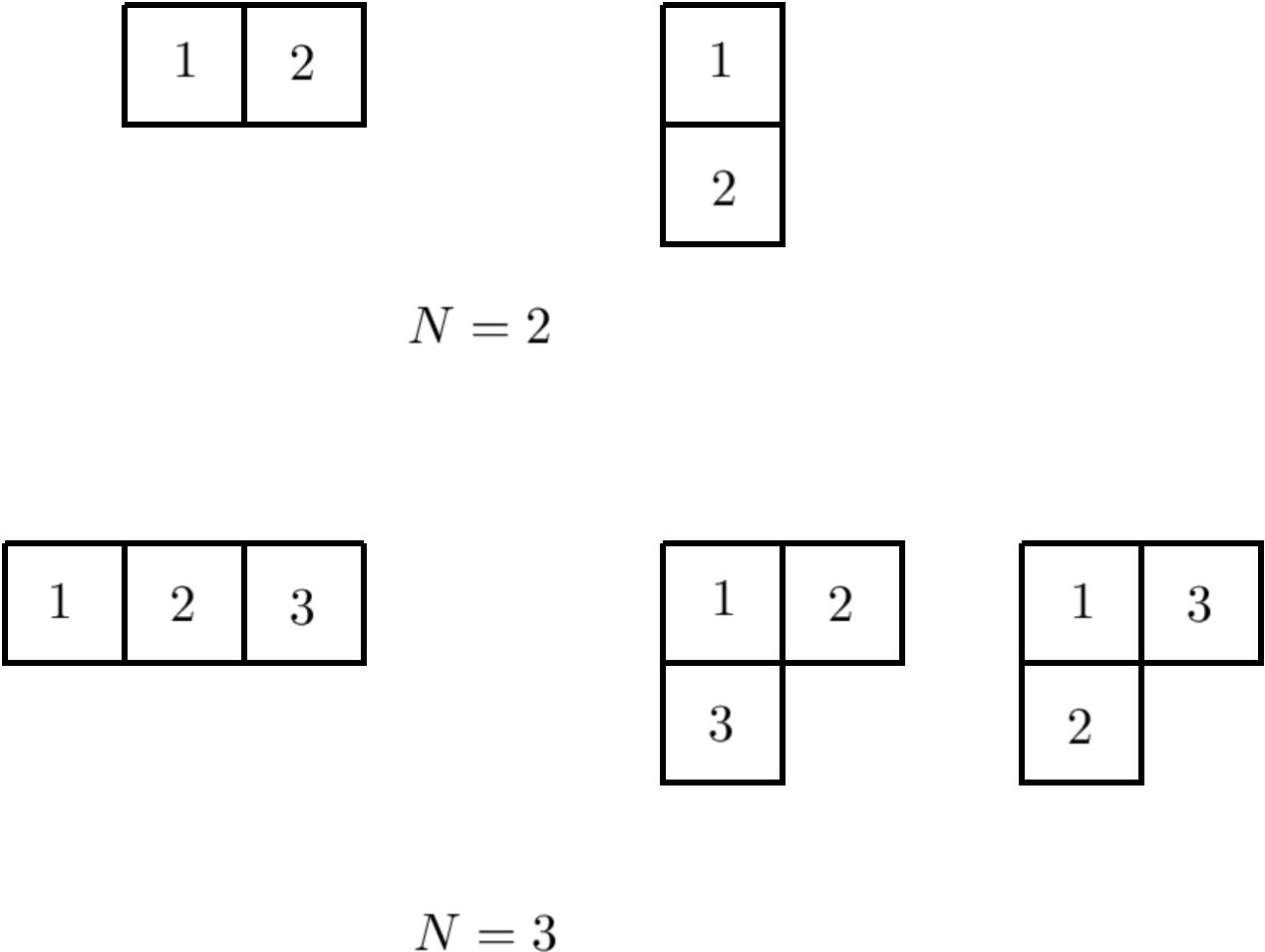}
     \caption{The standard Young diagrams for the $N=2$ and $N=3$. Each type of diagram gives rise to an IRR. So we have two one dimensional IRR's for $TL_2(x)$ and a one dimensional and a two dimensional IRR for $TL_3(x)$.}
      \label{fig:YoungDiagramsN2N3}
 \end{figure}

%%%%%%%%%%%%%%%%%%%%%%%%%%%%%%%%%%%%%%%%%%%
The standard Young diagrams with two rows carry irreducible representations (IRR's) of $TL_N(x)$ (See \cite{AnneMoore2008} for more details). The Young diagrams used in constructing these IRR's are made of two rows with the number of cells in the first row greater than equal to the number in the second row. The integers $\{1,2,\cdots, N\}$ fill the cells such that they are in ascending order along both the rows and each of the columns. For example the standard Young diagrams for IRR's of $TL_2(x)$ and $TL_3(x)$ are shown in Fig. \ref{fig:YoungDiagramsN2N3}.

%%%%%%%%%%%%%%%%%%%%%%%%%%%%%%%%%%%%%%%%%%%%%%%%%%%%%%%
%%%%%%%%%%%%%%%%%%%%%%%%%%%%%%%%%%%%%%%%%%%%%%%%%%%%%%%

The dimension of each IRR is obtained from the {\it Hook formula} that counts the number of standard Young diagrams of a given shape. This is given by
\begin{equation}
    \text{dim}(IRR) = \frac{N\, !}{\prod\limits_i~h_i},
\end{equation}
where $h_i$ is the {\it Hook length} for each cell indexed by $i$. This is equal to the number of cells to the right of cell $i$ plus the number of cells to the bottom of the $i$th cell plus one. From character theory \cite{hamermesh1989group}, the sum of the squares of the dimensions of the IRR's equals the $N$th Catalan number, $C_N = \frac{1}{N+1}\left(\begin{array}{c} 2N  \\ N \end{array} \right)$. Next we specify the action of $e_i$ on the standard Young diagrams denoted by $t$,
\begin{equation}\label{eq:TLactionYT}
    e_i.t = \begin{cases}
        C.t + \tilde{C}.s_i.t,~~s_i.t~\textrm{is standard}, \\
        C.t,~~~~~~~~~~~~~s_i.t~\textrm{is not standard}.
    \end{cases}
\end{equation}
Here $s_i$ denotes the permutation operator that exchanges the positions of the indices $i$ and $i+1$. The coefficients $C$ and $\tilde{C}$ are determined as 
\begin{equation}\label{eq:CcoefficientTL}
    C = \begin{cases}
        \frac{[d-1]}{[d]},~~ $i+1$~\textrm{is}~$d$~\textrm{steps North/East from}~ $i$ \\
        \frac{[d+1]}{[d]},~~ $i+1$~\textrm{is}~$d$~\textrm{steps South/West from}~ $i$,
    \end{cases}
\end{equation}
and 
\begin{equation}\label{eq:CtildecoefficientTL}
    \tilde{C} = \frac{\sqrt{[d-1][d+1]}}{[d]}.
\end{equation}
In these expressions the $[d]$ denotes polynomials in $x$. These are the $q$-numbers which appear in quantum group theory \cite{Majid1995FoundationsOQ,Biedenharn:1996vv}. We have 
\begin{equation}\label{eq:[d]terms}
    [d] = \frac{q^d-q^{-d}}{q-q^{-1}}=q^{d-1}+q^{d-3}+\cdots +q^{-(d-1)},~~x=q+q^{-1}.
\end{equation}
Some examples that will be relevant in this work are,
\begin{equation}
    [0]=0,~[1]=1,~[2]=x,~[3]=x^2-1.
\end{equation}
Using these rules the two IRR's of $TL_2(x)$ are one dimensional (See Fig. \ref{fig:YoungDiagramsN2N3}) and are given by, 
$$ e_1 = 0,~~e_1=x,$$
respectively. $TL_3(x)$ has two IRR's as well, one of which is one dimensional and the other is two dimensional (See Fig. \ref{fig:YoungDiagramsN2N3}). The generators in the one dimensional IRR are trivial,
$$ e_1 = 0,~~e_2=0,$$
while in the two dimensional IRR they are,
$$e_1=\left(\begin{array}{cc}
   0  & 0 \\
    0 & x
\end{array} \right),~~e_2=\left(\begin{array}{cc}
   \frac{x^2-1}{x}  & \frac{\sqrt{x^2-1}}{x} \\
    \frac{\sqrt{x^2-1}}{x} & \frac{1}{x}
\end{array} \right). $$

\begin{figure}[h]
     \centering
     \includegraphics[width=5cm]{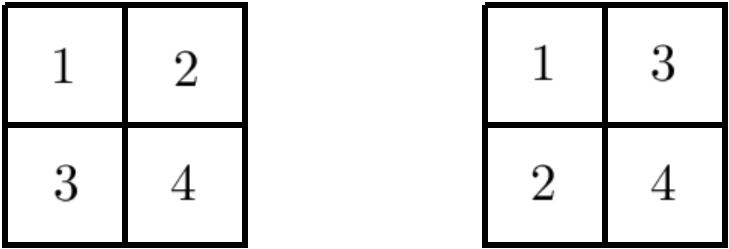}
     \caption{The standard Young diagrams of the $(2,2)$ type for $N=4$. }
      \label{fig:YoungDiagram22type}
 \end{figure}

\begin{figure}[h]
     \centering
     \includegraphics[width=13cm]{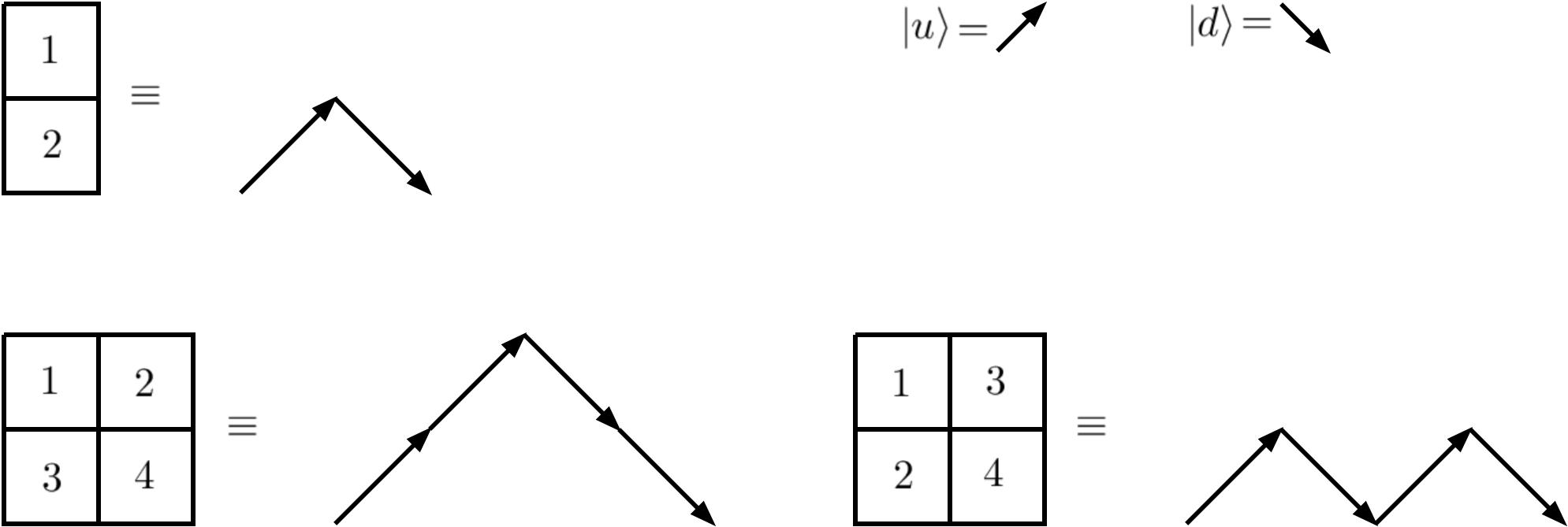}
     \caption{The mapping between the Dyck paths and standard Young diagrams of the $(N,N)$ type for $2N=2,4$. In terms of the `up' and `down' states the Young diagrams correspond to $\ket{ud}$, $\ket{uudd}$ and $\ket{udud}$ respectively.}
      \label{fig:YT2Dyck}
 \end{figure}

To establish the correspondence with the fusion basis of Fibonacci anyons we will be interested in the $(N,N)$ Young diagrams forming one IRR of $TL_{2N}(x)$. From the Hook formula it is easy to see that the dimension of this IRR is the $N$th Catalan number, $C_N$. In particular we will be interested in the $(2,2)$ IRR of $TL_4(x)$. This is a two dimensional IRR where the generators are given by 
\begin{equation}\label{eq:TL4generatorsYT}
    e_1=e_3=\left(\begin{array}{cc}
   0  & 0 \\
    0 & x
\end{array} \right),~~e_2=\left(\begin{array}{cc}
   \frac{x^2-1}{x}  & \frac{\sqrt{x^2-1}}{x} \\
    \frac{\sqrt{x^2-1}}{x} & \frac{1}{x}
\end{array} \right).
\end{equation}
They act on the standard diagrams in Fig. \ref{fig:YoungDiagram22type}.
%%%%%%%%%%%%%%%%%%%%%%%%%%%%%%%%%%%%%%%%%%%%%%%%%%%%%%%
%%%%%%%%%%%%%%%%%%%%%%%%%%%%%%%%%%%%%%%%%%%%%%%%%%%%%%%
The $N$th Catalan number also enumerates the number of Dyck paths on $2N$ steps. This suggests a one-to-one correspondence between the standard Young diagrams of the $(N,N)$ type and the length $2N$ Dyck paths. Indeed such a mapping exists \cite{grimaldi2012fibonacci}, and is built by associating an `up' (`down') step to each of the indices of the top (bottom) row of the standard Young diagram as shown {\it via} examples in Fig. \ref{fig:YT2Dyck}. 
%%%%%%%%%%%%%%%%%%%%%%%%%%%%%%%%%%%%%%%%%%%%%%%%%%%%%%%
%%%%%%%%%%%%%%%%%%%%%%%%%%%%%%%%%%%%%%%%%%%%%%%%%%%%%%%
This will be needed to make the correspondence between the anyon fusion basis states and the Dyck paths once we specify the final isomorphism.

%%%%%%%%%%%%%%%%%%%%%%%%%%%%%%%%%%%%%%%%%%%%%%%%%%%%%%%%%%%%%%%%%%
\subsection*{Representation on the Fibonacci anyon fusion basis}
\label{subsec:Fibrep}
\begin{figure}[h]
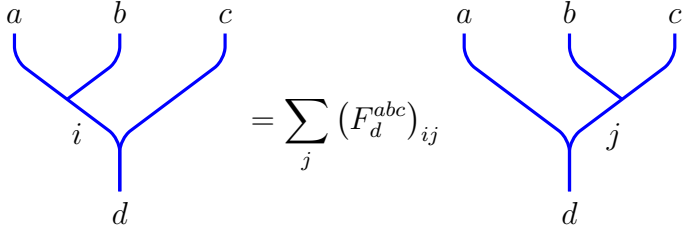

		\centering
		\begin{align*}
		\tikz[baseline=-7ex, scale = 0.7]{
			\def\a{0}
			\def\b{0}
			\draw[very thick, blue, rounded corners=5] (\a, \b) -- (\a, \b-0.5) -- (2+\a, \b-2) -- (2+\a, \b -3);
			\draw[very thick, blue, rounded corners=5] (\a + 2, \b) -- (\a + 2, \b-0.5) -- (1+\a, \b-1.25);
			\draw[very thick, blue, rounded corners=5] (\a + 4, \b) -- (\a + 4, \b-0.5) -- (2+\a, \b-2) -- (2+\a, \b -3); 
			%\draw[thick, blue] (2+\a, \b-2) -- (2+\a, \b -3);
			\node[anchor=south] at (\a, \b) {$a$};
			\node[anchor=south] at (\a + 2, \b) {$b$};
			\node[anchor=south] at (\a + 4, \b) {$c$};
			\node[anchor=north east] at (\a + 1.5, \b - 1.5) {$i$};
			\node[anchor=north] at (\a + 2, \b - 3) {$d$};
		} = \sum_{j}\left(F_{d}^{abc}\right)_{ij}
		\tikz[baseline=-7ex, scale = 0.7]{
			\def\a{0}
			\def\b{0}
			\draw[very thick, blue, rounded corners=5] (\a, \b) -- (\a, \b-0.5) -- (2+\a, \b-2) -- (2+\a, \b -3);
			\draw[very thick, blue, rounded corners=5] (\a + 2, \b) -- (\a + 2, \b-0.5) -- (3+\a, \b-1.25);
			\draw[very thick, blue, rounded corners=5] (\a + 4, \b) -- (\a + 4, \b-0.5) -- (2+\a, \b-2) -- (2+\a, \b -3); 
			%\draw[thick, blue] (2+\a, \b-2) -- (2+\a, \b -3);
			\node[anchor=south] at (\a, \b) {$a$};
			\node[anchor=south] at (\a + 2, \b) {$b$};
			\node[anchor=south] at (\a + 4, \b) {$c$};
			\node[anchor=north west] at (\a + 2.5, \b - 1.5) {$j$};
			\node[anchor=north] at (\a + 2, \b - 3) {$d$};
		}
		\end{align*}
	\caption{The $F$-move as a sliding of the $b$ line from $a$ to $c$.} 
 \label{fig:Fmove}
	\end{figure}

\begin{figure}[h]
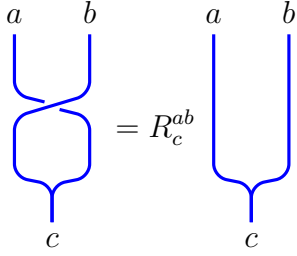

	\centering
	\begin{align*}
		\tikz[baseline=3.5ex]{
			\def\a{0}
			\def\b{0}
			\draw[very thick, blue, rounded corners=5] (0.5+\a, 2+\b) -- (0.5+\a, 1.2+\b) -- (\a - 0.5, 0.9+\b) -- (-0.5+\a, \b + 0.1) -- (\a, \b) -- (\a, \b - 0.5);
			\draw[very thick, blue, rounded corners=5] (-0.5+\a, 2+\b) -- (-0.5+\a, 1.2+\b) -- (\a-0.1, 1.1+\b);
			\draw[very thick, blue, rounded corners=5] (0.1+\a, 1+\b) -- (\a + 0.5, 0.9+\b) -- (0.5+\a, \b + 0.1) -- (\a, \b) -- (\a, \b - 0.5);
			\node[anchor=south] at (\a - 0.5, 2 + \b) {$a$};
			\node[anchor=south] at (\a + 0.5, 2 + \b) {$b$};
			\node[anchor=north] at (\a, \b - 0.5) {$c$};
		} = R_{c}^{ab}
		\tikz[baseline=3.5ex]{
			\def\a{0}
			\def\b{0}
			\draw[very thick, blue, rounded corners=5] (0.5+\a, 2+\b) -- (0.5+\a, \b + 0.1) -- (\a, \b) -- (\a, \b - 0.5);
			\draw[very thick, blue, rounded corners=5] (-0.5+\a, 2+\b) -- (-0.5+\a, \b + 0.1) -- (\a, \b) -- (\a, \b - 0.5);
			\node[anchor=south] at (\a - 0.5, 2 + \b) {$a$};
			\node[anchor=south] at (\a + 0.5, 2 + \b) {$b$};
			\node[anchor=north] at (\a, \b - 0.5) {$c$};
		} 
	\end{align*}
	\caption{The $R$-move untwists fusing anyon lines. 
		%The multiplicities are suppressed. 
		Note that the anyon line $b$ crosses over $a$. An undercrossing would correspond to  using $R^{-1}$ instead of $R$.}
	\label{fig:Rmove}
\end{figure}

%%%%%%%%%%%%%%%%%%%%%%%%%%%%%%%%%%%%%%%%%%%%%%%%%%%%%%%%%%%%%%%%%%
Next we come to the last ingredient in the construction of the isomorphism, the Jones representation on the fusion space of Fibonacci anyons. From the categorical perspective this is an example of a unitary braided fusion category made of the object labels $\{1, \tau\}$, with $1$ being the vacuum or trivial particle and $\tau$ being the Fibonacci anyon. To define this category we begin with the {\it fusion rules},
\begin{equation}\label{eq:fusionFib}
    1\times\tau=\tau\times 1= \tau,~~\tau\times\tau = 1 + \tau.
\end{equation}
These anyons can also be seen as IRR's of the quantum group, $SU(2)_3$ and the fusion rules are the composition of the different IRR's into a direct sum. Along with the fusion data we require the $F$- and $R$-matrices to determine the Fibonacci anyon category. Pictorially these matrices represent the $F$- and $R$-moves as shown in Figs. \ref{fig:Fmove}, \ref{fig:Rmove}.
%%%%%%%%%%%%%%%%%%%%%%%%%%%%%%%%%%%%%%%%%%%%%%%%%%%
%%%%%%%%%%%%%%%%%%%%%%%%%%%%%%%%%%%%%%%%%%%%%%%%%%%
%%%%%%%%%%%%%%%%%%%%%%%%%%%%%%%%%%%%%%%%%%%%%%%%%%%
%%%%%%%%%%%%%%%%%%%%%%%%%%%%%%%%%%%%%%%%%%%%%%%%%%%

\begin{figure}[h]
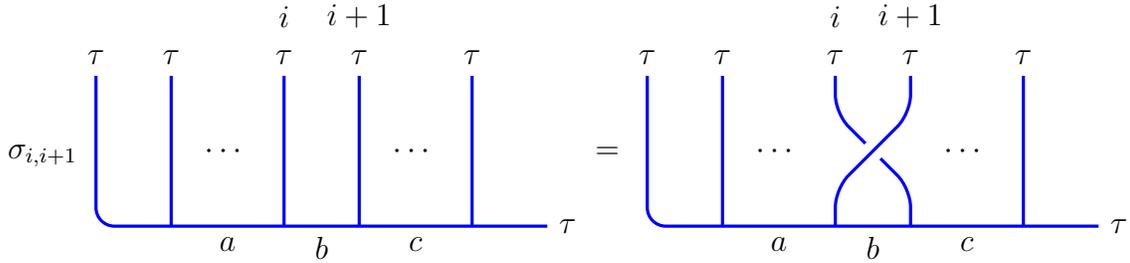

	\centering
	\begin{align*}
		\sigma_{i, i+1}\tikz[baseline=5ex, scale=1]{
			\def\a{0}
			\def\b{0}
			\node[anchor=south] at (\a + 2.5, \b + 2.5) {$i$};
			\node[anchor=south] at (\a + 3.5, \b + 2.5) {$i+1$};
			\draw[very thick, blue, rounded corners=7] (\a, \b+2) node[anchor=south, black] {$\tau$} -- (\a, \b) -- (\a + 6, \b) node[anchor=west, black] {$\tau$};
			\draw[very thick, blue] (\a+1, \b) -- (\a+1, \b+2) node[anchor=south, black] {$\tau$};
			\node[anchor=west] at (\a+1.3, \b+1) {$\cdots$};
			\draw[very thick, blue] (\a+2.5, \b) -- (\a+2.5, \b+2) node[anchor=south, black] {$\tau$};
			\draw[very thick, blue] (\a+3.5, \b) -- (\a+3.5, \b+2) node[anchor=south, black] {$\tau$};
			\node[anchor=west] at (\a+3.8, \b+1) {$\cdots$};
			\draw[very thick, blue] (\a+5, \b) -- (\a+5, \b+2) node[anchor=south, black] {$\tau$};
			\node[anchor=north] at (\a + 1.75, \b) {$a$};
			\node[anchor=north] at (\a + 3, \b) {$b$};
			\node[anchor=north] at (\a + 4.25, \b) {$c$};
		}=
		\tikz[scale=1, baseline=5ex]{
			\def\a{0}
			\def\b{0}
			\node[anchor=south] at (\a + 2.5, \b + 2.5) {$i$};
			\node[anchor=south] at (\a + 3.5, \b + 2.5) {$i+1$};
			\draw[very thick, blue, rounded corners=7] (\a, \b+2) node[anchor=south, black] {$\tau$} -- (\a, \b) -- (\a + 6, \b) node[anchor=west, black] {$\tau$};
			\draw[very thick, blue] (\a+1, \b) -- (\a+1, \b+2) node[anchor=south, black] {$\tau$};
			\node[anchor=west] at (\a+1.3, \b+1) {$\cdots$};
			\def\a{1.5}
			\def\b{2}
			\draw[very thick, blue, rounded corners=7] (\a + 1, \b) node[anchor=south, black] {$\tau$} -- (\a + 1, \b - 0.5) -- (\a + 1.4, \b - 0.9);
			\draw[very thick, blue, rounded corners=7] (\a + 1.6, \b - 1.1) -- (\a + 2, \b - 1.5) -- (\a + 2, \b - 2);
			\draw[very thick, blue, rounded corners=7] (\a + 2, \b) node[anchor=south, black] {$\tau$} -- (\a + 2, \b - 0.5) -- (\a + 1, \b - 1.5) -- (\a + 1, \b - 2);
			\def\a{0}
			\def\b{0}
			\node[anchor=west] at (\a+3.8, \b+1) {$\cdots$};
			\draw[very thick, blue] (\a+5, \b) -- (\a+5, \b+2) node[anchor=south, black] {$\tau$};
			\node[anchor=north] at (\a + 1.75, \b) {$a$};
			\node[anchor=north] at (\a + 3, \b) {$b$};
			\node[anchor=north] at (\a + 4.25, \b) {$c$};
		}
	\end{align*}
	\caption{Action of $\sigma_i$ on the anyon fusion basis.}
	\label{fig:sigmaiactionfusion}
\end{figure}

This graphical notation facilitates the construction of the braid group, and consequently the Temperley-Lieb algebra, on the fusion spaces. This follows from the action of the braid generators on the fusion basis elements as shown in Fig. \ref{fig:sigmaiactionfusion}.
%%%%%%%%%%%%%%%%%%%%%%%%%%%%%%%%%%%%%%%%%%%%%%%%%%%%%%
%%%%%%%%%%%%%%%%%%%%%%%%%%%%%%%%%%%%%%%%%%%%%%%%%%%%%%%%%%%%%%%%%%%%%%%%%%%
The $F$- and $R$-moves are then applied to these diagrams to reduce it to a linear combination of fusion basis elements (See \cite{kauffmanlomonaco,Jana2022TopologicalQC} for more detail).

In this work we are interested in the representations of $\mathcal{B}_3$ that act on the fusion basis obtained by fusing three $\tau$ anyons. The resulting representation is two dimensional, acting on the basis $\{\ket{1}, \ket{\tau} \}$ and is given by,
\begin{equation}\label{eq:FibBraidreps}
    \sigma_1 = R,~~ \sigma_2 = FRF,
\end{equation}
with 
\begin{equation}\label{eq:FRmatricesFib}
    F= \left(\begin{array}{cc}
       \phi^{-1}  & \phi^{-\frac{1}{2}}  \\
       \phi^{-\frac{1}{2}}  & -\phi^{-1} 
    \end{array} \right),~~ R = \left(\begin{array}{cc}
       e^{\frac{4\pi\mathrm{i}}{5}}  & 0 \\
       0  &  e^{-\frac{3\pi\mathrm{i}}{5}}
    \end{array} \right), 
\end{equation}
where $\phi=\frac{1+\sqrt{5}}{2}$ is the Golden ratio. Along with the one dimensional representation of $\mathcal{B}_2$, $\sigma_1 = R^{\tau\tau}_\tau=e^{-\frac{3\pi\mathrm{i}}{5}}$, the representation of $\mathcal{B}_3$ are irreducible. The braid groups, $\mathcal{B}_N$ for $N>3$, obtained by fusing more $\tau$ anyons are built using these IRR's. As mentioned earlier we will focus on the two dimensional IRR as we will use that to construct the desired isomorphism.

\begin{figure}[h]
     \centering
     \includegraphics[width=12cm]{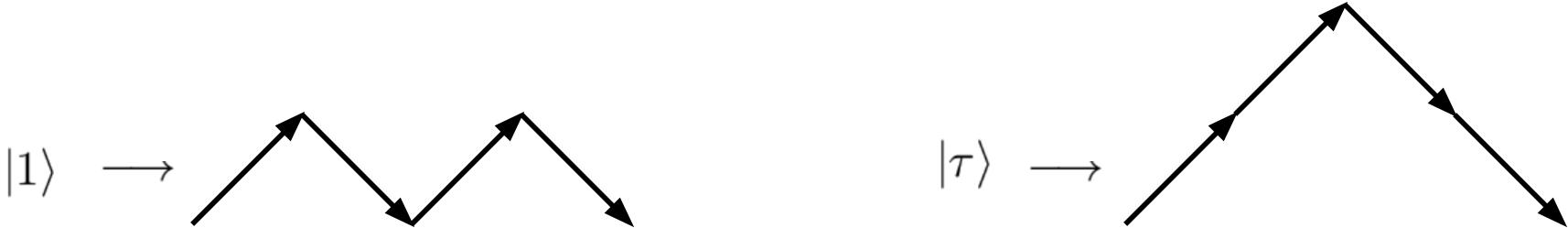}
     \caption{Correspondence between the fusion basis of three Fibonacci anyons and length 4 Dyck paths. }
      \label{fig:fusion3basis2DyckPath}
 \end{figure}

The braid generators in \eqref{eq:FibBraidreps} can be put in the Jones form \eqref{eq:sigmaJonesrep}, to reveal the representations of the Temperley-Lieb generators on the fusion space. We find the $TL_3(x)$ generators to be,
\begin{equation}
    \tilde{e}_1 = \left(\begin{array}{cc}
       \phi  & 0 \\
       0  & 0
    \end{array} \right),~~ \tilde{e_2} = \left(\begin{array}{cc}
        \phi^{-1} & \phi^{-\frac{1}{2}} \\
        \phi^{-\frac{1}{2}} & 1
    \end{array} \right).
\end{equation}
By setting $x=\phi$ and using the Pauli $X$ matrix we can rotate these matrices to match the generators of $TL_4(x)$ in \eqref{eq:TL4generatorsYT}. This isomorphism between the two IRR's lead to our desired correspondence between the length 4 Dyck states and the Fibonacci fusion basis,
\begin{equation}\label{eq:Fibstates2Dyckpaths}
    \ket{1} \rightarrow \ket{udud},~~ \ket{\tau} \rightarrow \ket{uudd}.
\end{equation}
Pictorially this is depicted in Fig. \ref{fig:fusion3basis2DyckPath}.
%%%%%%%%%%%%%%%%%%%%%%%%%%%%%%%%%%%%%%%%%%%%%%%%%%%%%%%
%%%%%%%%%%%%%%%%%%%%%%%%%%%%%%%%%%%%%%%%%%%%%%%%%%%%%%%

With this mapping in place we can write down the entire fusion basis of an arbitrary number of fusing $\tau$ anyons in terms of the length 4 Dyck paths. The resulting states are sequences of $\ket{udud}$ and $\ket{uudd}$ such that consecutive $\ket{udud}$'s do not appear. This mimics the construction of the Fibonacci sequences from the anyons $1$ and $\tau$. 

%%%%%%%%%%%%%%%%%%%%%%%%%%%%%%%%%%%%%%%%%%
\subsection*{Braid group on Dyck paths}
\label{subsec:braidgroupDyckpaths}
%%%%%%%%%%%%%%%%%%%%%%%%%%%%%%%%%%%%%%%%%%
We can write down explicit expressions for the braid generators acting on the space of Dyck paths. For example the generators of the Temperley-Lieb generators, $e_1$, $e_2$ and $e_3$ in \eqref{eq:TL4generatorsYT} as operators acting on the two length 4 Dyck paths are,
\begin{eqnarray}\label{eq:TL4generatorsDyck}
    e_1=e_3 & = & x~\left(\frac{1+Z_1}{2}\right)\left(\frac{1-Z_2}{2}\right)\left(\frac{1+Z_3}{2}\right)\left(\frac{1-Z_4}{2}\right), \nonumber \\
    e_2 & = & \frac{x^2-1}{x}~\left(\frac{1+Z_1}{2}\right)\left(\frac{1+Z_2}{2}\right)\left(\frac{1-Z_3}{2}\right)\left(\frac{1-Z_4}{2}\right) \nonumber \\
    & + & \frac{1}{x}~\left(\frac{1+Z_1}{2}\right)\left(\frac{1-Z_2}{2}\right)\left(\frac{1+Z_3}{2}\right)\left(\frac{1-Z_4}{2}\right) \nonumber \\
    & + & \frac{\sqrt{x^2-1}}{x}~\left(\frac{1+Z_1}{2}\right)\left(\frac{X_2X_3 + Y_2Y_3}{2}\right)\left(\frac{1-Z_4}{2}\right),
\end{eqnarray}
where the suffixes on the Pauli $X$ and $Z$ operators denote the four sites of the length 4 chain. It is easily verified that these generators satisfy the Temperley-Lieb relations in \eqref{eq:TLrelations}. Following this the braid generators are constructed using the Jones representation, \eqref{eq:sigmaJonesrep}. The braid group on an arbitrary number of strands, $\mathcal{B}_N$, can be constructed in a graphical manner similar to the one used above. In this case we split the Temperley-Lieb generators into three, those acting on the left ($e_1$, $e_2$), the bulk ($e_i$, $i\in\{3,\cdots, N-2\}$) and the right ($e_{N-1}$). For completion we write down these expressions in App. \ref{app:BNDyckPaths}.
%%%%%%%%%%%%%%%%%%%%%%%%%%%%%%%%%%%%%%%%%%%%%%%%%%%%%%%%%%%%%%%%%%%%%%%%%%%%%%
\section{The Spin Chain from Fredkin Moves}
\label{sec:spinchain}
\begin{figure}[h]
     \centering
     \includegraphics[width=6cm]{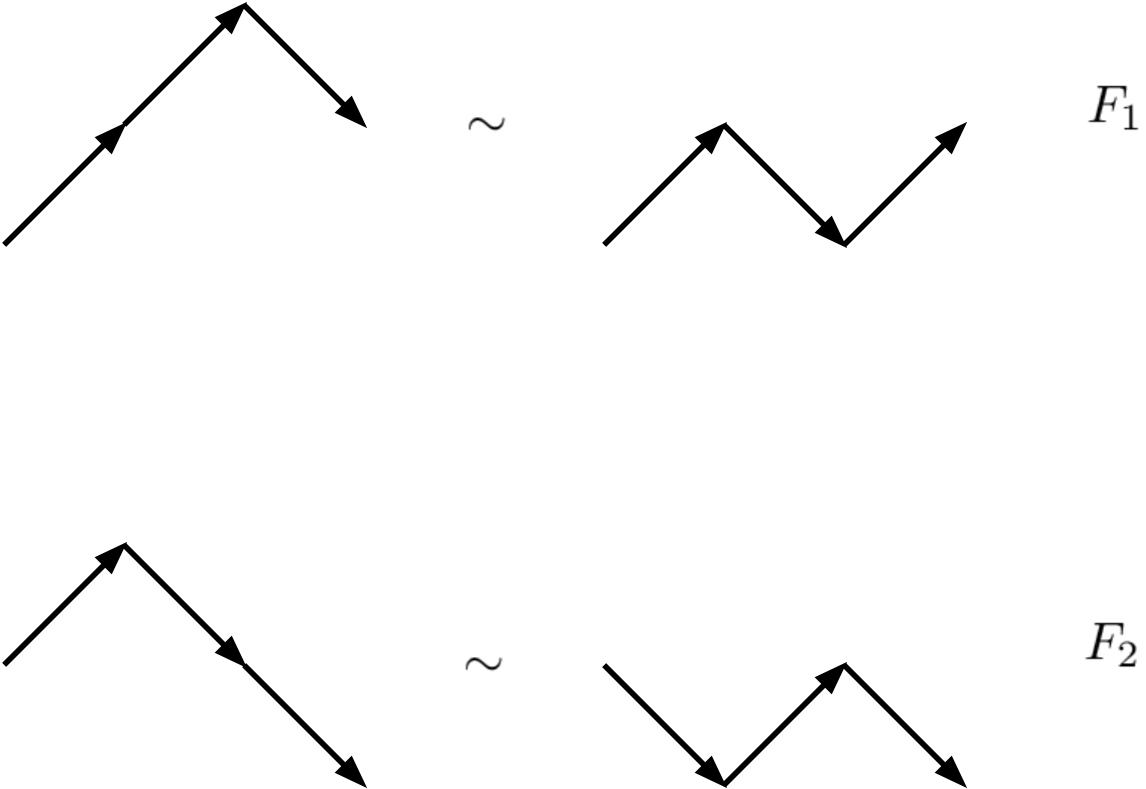}
     \caption{The local Fredkin moves, $F_1$, $F_2$, acting on three neighboring physical qubits. The Dyck paths are generated by identifying the configurations on either side of the $\sim$.}
      \label{fig:fredkinmoves}
 \end{figure}

\begin{figure}[h]
     \centering
     \includegraphics[width=10cm]{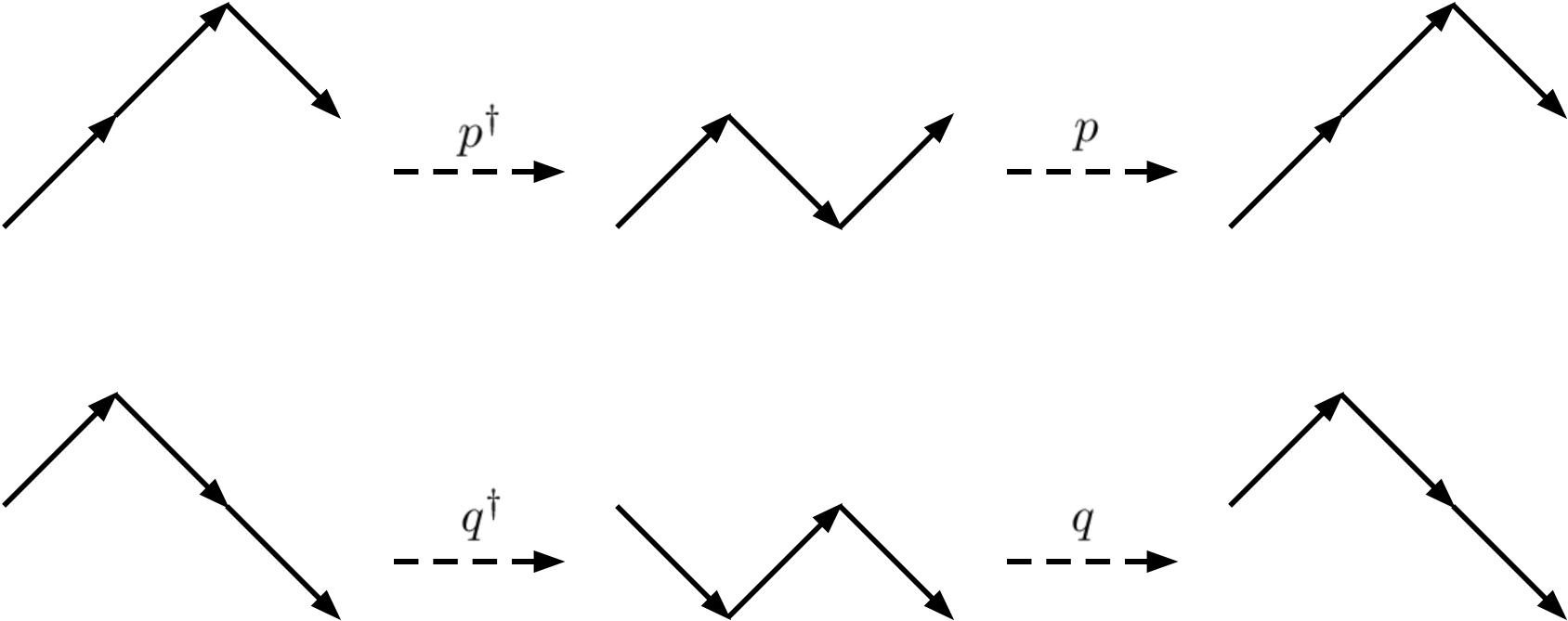}
     \caption{The Fredkin operators in \eqref{eq:fredkinoperators} in graphical form. This is useful for doing computations on the space of Dyck paths.}
      \label{fig:fredkinoperators}
 \end{figure}

\begin{figure}[h]
     \centering
     \includegraphics[width=10cm]{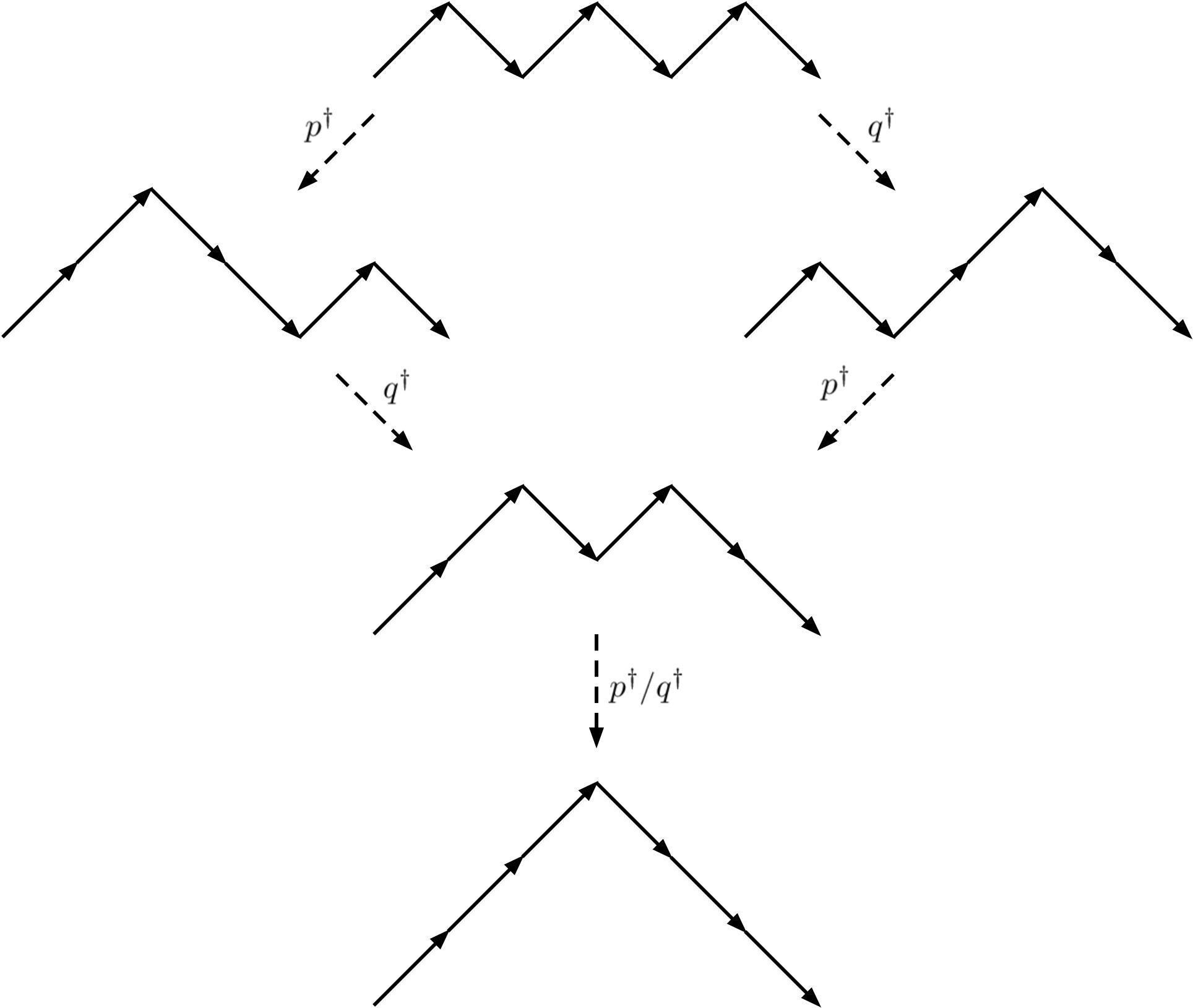}
     \caption{The five $N=6$ Dyck paths generated by application of the $p$ and $q$ operators.}
      \label{fig:DyckpathsN6}
 \end{figure}

%%%%%%%%%%%%%%%%%%%%%%%%%%%%%%%%%%%%%%%%%%%%%%%%%%%%%%%%%%%%%%%%%%%%%%%%%%%%%%
Now we move on to the construction of the spin chain that includes the Fibonacci sequences made out of the two length 4 Dyck paths (See Fig. 
\ref{fig:fusion3basis2DyckPath}) as a degenerate set of eigenstates. We want the Hamiltonian to be made out of local terms and these are precisely provided by the {\it Fredkin moves} (See Fig. \ref{fig:fredkinmoves}), which are local moves that map between the different Dyck paths of length $2N$ \cite{Salberger2016FredkinSC}. 
%%%%%%%%%%%%%%%%%%%%%%%%%%%%%%%%%%%%%%%%%%%%%%%%%%%%%%%
%%%%%%%%%%%%%%%%%%%%%%%%%%%%%%%%%%%%%%%%%%%%%%%%%%%%%%%
It is a useful exercise to see that all the Dyck paths of length $2N$ can be generated from the state $\ket{udud\cdots ud}$ by repeated use of the Fredkin moves, $F_1$ and $F_2$, in Fig. \ref{fig:fredkinmoves}. To do this we first create the nilpotent operators from the Fredkin moves, $F_1$ and $F_2$,
\begin{eqnarray}\label{eq:fredkinoperators}
    p_j^\dag & = & \ket{udu}\bra{uud},~~p_j  =  \ket{uud}\bra{udu}, \nonumber \\
    q_j^\dag & = & \ket{dud}\bra{udd},~~q_j  =  \ket{udd}\bra{dud}.
\end{eqnarray}
These operators act on the consecutive sites, $\{j, j+1, j+2\}$. It is convenient to view this in graphical form, Fig. \ref{fig:fredkinoperators}.
%%%%%%%%%%%%%%%%%%%%%%%%%%%%%%%%%%%%%%%%%%%%%%%%%%%%%%%
%%%%%%%%%%%%%%%%%%%%%%%%%%%%%%%%%%%%%%%%%%%%%%%%%%%%%%%
The paths generated this way is shown for $N=6$ in Fig. \ref{fig:DyckpathsN6}.
%%%%%%%%%%%%%%%%%%%%%%%%%%%%%%%%%%%%%%%%%%%%%%%%%%%%%%%
%%%%%%%%%%%%%%%%%%%%%%%%%%%%%%%%%%%%%%%%%%%%%%%%%%%%%%%

The desired spin chain is constructed on a chain with $4N$ sites with the local Hilbert space on each site being spanned by $\{\ket{u}, \ket{d}\}$.
The Fibonacci sequences that we want as eigenstates are composed of just the two length 4 Dyck paths, Fig. \ref{fig:fusion3basis2DyckPath}. The order of the resulting degenerate set is the Fibonacci number which is much lesser than the Catalan number which in turn is lesser than the dimension of the full Hilbert space, $2^{4N}$. The spin chain is made of two local terms. 
\begin{enumerate}
    \item The first term, $H_{4j-3}^{(4)}$ acts on four consecutive sites, such that the two length 4 Dyck paths form a degenerate set. The remaining fourteen states are separated from this set in energy. The full system with a sum of just these 4-site Hamiltonians will include all combinations of the two length 4 Dyck paths as a degenerate set of eigenstates corresponding to $2^N$ states. However two consecutive states of the form $\ket{udud}$ are forbidden (See Fig. \ref{fig:fusion3basis2DyckPath}). To remove this we will include an interaction term between two disjoint and consecutive 4-sites.
    \item The interaction term $H^{(4)}_{4j-1}$\footnote{This term is not unique. A 6-site term, $ H^{(6)}_{4j-2} = q_{4j-2}^\dag q_{4j-2}p_{4j+1}^\dag p_{4j+1}$ will also project out the forbidden sequence. For simplicity we will stick to the 4-site term for the rest of this paper. }, acts on four physical qubits, $\{4j-1,\cdots, 4j+2\}$ common to the two consecutive sets . This term will lift the forbidden sequence out of the desired degenerate set of eigenstates reducing the number of such states to precisely the Fibonacci number.
\end{enumerate}

The total Hamiltonian is given by,
\begin{equation}\label{eq:FredkinHamiltonian}
    H = \sum\limits_{j=1}^N~H_{4j-3}^{(4)} + \lambda~\sum\limits_{j=1}^{N-1}~H^{(4)}_{4j-1},
\end{equation}
with 
\begin{eqnarray}\label{eq:4sitetermFredkinHamiltonian}
    H_{4j-3}^{(4)} & = & \alpha_1~\left(p_{4j-3}^\dag p_{4j-3} + p_{4j-3} p_{4j-3}^\dag \right) + \alpha_2~\left(p_{4j-2}^\dag p_{4j-2} + p_{4j-2} p_{4j-2}^\dag \right) \nonumber \\
    & + & \beta_1~\left(q_{4j-3}^\dag q_{4j-3} + q_{4j-3} q_{4j-3}^\dag \right) + \beta_2~\left(q_{4j-2}^\dag q_{4j-2} + q_{4j-2} q_{4j-2}^\dag \right),
\end{eqnarray}
and the `quadratic' interaction term,
\begin{equation}\label{eq:4sitetermFredkinHamiltonian}
  H^{(4)}_{4j-1} = p^\dag_{4j-1}p_{4j-1}q^\dag_{4j}q_{4j}.  
\end{equation}
The parameters $\alpha_1$, $\alpha_2$, $\beta_1$, $\beta_2$ and $\lambda$ are real for the Hamiltonian to be hermitian. It is immediately clear that the terms making up this Hamiltonian are projectors to particular configurations on either three sites or four sites. Thus the Hamiltonian is diagonal in the local $\{\ket{u}, \ket{d}\}$ basis with,
\begin{eqnarray}\label{eq:4siteHamiltonianZ}
   H_{4j-3}^{(4)} & = & \alpha_1~\left(\frac{1+Z_{4j-3}}{2}\right)\left(\frac{1-Z_{4j-2}Z_{4j-1}}{2}\right) + \alpha_2~\left(\frac{1+Z_{4j-2}}{2}\right)\left(\frac{1-Z_{4j-1}Z_{4j}}{2}\right) \nonumber \\
   & + & \beta_1~\left(\frac{1-Z_{4j-3}Z_{4j-2}}{2}\right)\left(\frac{1-Z_{4j-1}}{2}\right) + \beta_2~\left(\frac{1-Z_{4j-2}Z_{4j-1}}{2}\right)\left(\frac{1-Z_{4j}}{2}\right), \nonumber \\
\end{eqnarray}
and
\begin{equation}\label{eq:4siteHamiltonianZ}
 H^{(4)}_{4j-1} = \left(\frac{1+Z_{4j-1}}{2}\right)\left(\frac{1-Z_{4j}}{2}\right)\left(\frac{1+Z_{4j+1}}{2}\right)\left(\frac{1-Z_{4j+2}}{2}\right). 
\end{equation}
The diagonal form of the Hamiltonian in this basis implies that all the eigenstates are just product states. While this may appear trivial it is still necessary to show that a set of local terms can project out the desired set of eigenstates. It is also a sufficient condition to show that the length 4 Dyck paths form a degenerate set with Fibonacci order and are separated from the rest of the spectrum by an energy gap. To do this we begin by looking at the spectrum of the 4-site Hamiltonian (Table \ref{tab:spectrum4site}) which shows the two length 4 Dyck paths are degenerate and separated from the other states.

\begin{table}[h!]
  \begin{center}
    \begin{tabular}{|c|c|c|} 
     \hline
      Eigenstates & Eigenvalue & Degeneracy  \\
     \hline
      $\ket{uuuu}$, $\ket{dddd}$, $\ket{dddu}$, $\ket{dduu}$, $\ket{duuu}$  & 0 & 5\\
      $\ket{uduu}$ & $\alpha_1$  & 1\\
      $\ket{uuud}$, $\ket{duud}$ & $\alpha_2$ & 2\\
      $\ket{uddd}$, $\ket{uddu}$ & $\beta_1$ & 2 \\
      $\ket{ddud}$ & $\beta_2$ & 1 \\
      $\ket{uudu}$ & $\alpha_1+\alpha_2$ & 1 \\
      $\ket{dudd}$ & $\beta_1+\beta_2$ & 1 \\
      $\ket{dudu}$ & $\alpha_2+\beta_1$ & 1 \\
      \color{red}{$\ket{udud}$, $\ket{uudd}$} & \color{red}{$\alpha_1+\beta_2$} & \color{red}{2} \\
      \hline
\end{tabular}
    \caption{The spectrum of $H_{4j-3}^{(4)}$ for a particular $j$. The two length 4 Dyck paths are shown in \textcolor{red}{red} (last row). }
    \label{tab:spectrum4site}
  \end{center}
\end{table}  
The fusion basis of the Fibonacci anyons (See Fig. \ref{fig:fusion3basis2DyckPath}) is recovered in the eigenspace with value $N\left(\alpha_1+\beta_2\right)$ and dimension being the $N$th Fibonacci number. This corresponds to the fusion of $N$ Fibonacci anyons. They are separated from the rest of the spectrum by non-zero multiples of $\lambda$. Next we will show that this eigenspace can be gapped from the rest of the spectrum.

%%%%%%%%%%%%%%%%%%%%%%%%%%%%%%%%%%%%%%%%%%%%%%%%%%%%%%%%%%%%%%%%%%%%%%%%%%%%%%
\subsection{Energy Gap}
\label{sec:energygap}
%%%%%%%%%%%%%%%%%%%%%%%%%%%%%%%%%%%%%%%%%%%%%%%%%%%%%%%%%%%%%%%%%%%%%%%%%%%%%%
A system is gapped if there is a constant difference between one of the extreme eigenvalues and the rest of the spectrum as the size of the system increases. To establish the gap we first tune the parameters on a single set of 4 sites. We separate the energy level $\alpha_{1}+\beta_{2}$ from the rest of the spectrum to the highest extreme\footnote{This can be made the lowest eigenvalue by simply reversing signs.} by imposing the following condition
\begin{align*}
    \alpha_{1}+\beta_{2}> \max\{\alpha_{1}, \alpha_{2}, \beta_{1}, \beta_{2}, \alpha_{1}+\alpha_{2}, \beta_{1}+\beta_{2}, \alpha_{2}+\beta_{1}\}.
\end{align*}
This condition can be equivalently written as
\begin{align*}
    \alpha_{1}, \beta_{2}  > 0,\;\;
    \beta_{2}  > \alpha_{2},\;\;
    \alpha_{1}  > \beta_{1}.\;\;
    %\alpha_{1} + \beta_{2}  > \max\{\alpha_{2}, \beta_{1}\}.
\end{align*}
Combining the above inequalities we obtain
\begin{align}\label{eqn: energy gap}
    \alpha_{1} > \beta_{1} > 0,\;\;\beta_{2} > \alpha_{2}>0
    %\alpha_{1}+\beta_{2} > \max\{0, \alpha_{2}, \beta_{1}, \alpha_{2} + \beta_{1}\}.
\end{align}

On the extended chain the gap is determined by the interacting part of the Hamiltonian (i.e., $H_{4j - 1}^{(4)}$) controlled by $\lambda$. The energy level of an arbitrary state on the full chain takes the form
\begin{align*}
    k\lambda + \sum_{i=1}^{N}a_{i},
\end{align*}
where $0\leq k\leq N-1$ and $a_{i}\in \{0, \alpha_{1}, \alpha_{2}, \beta_{1}, \beta_{2}, \alpha_{1}+\alpha_{2}, \beta_{1}+\beta_{2}, \alpha_1 + \beta_2\}$ for all $i$. The maximum contribution from the interaction term is $(N-1)\lambda$ and there are sixteen such states determined by the boundary configurations $\{\ket{uu}, \ket{ud}, \ket{du}, \ket{dd} \}$ on the first two and the last two sites. The `bulk' spanning the sites $\{3, \cdots, 4N-2 \} $ are filled by $\ket{udud\cdots ud}$. These states have the eigenvalue $N\left(\alpha_1 + \beta_2 \right) + (N-1)\lambda$. Since we require $N\left(\alpha_1 + \beta_2 \right)$ to be the highest eigenvalue, $\lambda$ is taken to be negative. Thus the system can have an energy gap by tuning the parameters to satisfy the inequality \eqref{eqn: energy gap} and this is
\begin{align}\label{eqn:the energy gap}
     \alpha_{1}+\beta_{2} - \max\{\alpha_{1}, \alpha_{2}, \beta_{1}, \beta_{2}, \alpha_{1}+\alpha_{2}, \beta_{1}+\beta_{2}, \alpha_{2}+\beta_{1}\}.
\end{align}

%The energy gap is determined by $\lambda$ provided it satisfies, 
%\begin{align}\label{eqn: energy gap lambda}
 %   \lambda < \frac{N}{N-1}[\alpha_{1}+\beta_{2} - \max\{\alpha_{1}, \alpha_{2}, \beta_{1}, \beta_{2}, \alpha_{1}+\alpha_{2}, \beta_{1}+\beta_{2}, \alpha_{2}+\beta_{1}\}].
%\end{align}
 
%%%%%%%%%%%%%%%%%%%%%%%%%%%%%%%%%%%%%%%%%%%%%%%%%%%%%%%%%%%%%%%%%%%%%%%%%%%%%%
\section{Stability}
\label{sec:stability}
%%%%%%%%%%%%%%%%%%%%%%%%%%%%%%%%%%%%%%%%%%%%%%%%%%%%%%%%%%%%%%%%%%%%%%%%%%%%%%
The Hamiltonian in \eqref{eq:FredkinHamiltonian} is robust to diagonal perturbations that are smaller than the gap. These diagonal terms generate a continuous set of symmetries and protect the system as long as the perturbations stay within the gap. However we would like to go further and check if the gap is stable to perturbations that do not preserve these symmetries, in particular we look at random, off-diagonal noise. We do this by analysing Hamiltonians of the form $H+M$, where $H$ is the system Hamiltonian in \eqref{eq:FredkinHamiltonian} and $M$ is a hermitian noise matrix. We consider two situations. In the first case each row of $M$ contains only a few random variables and here we show that the circle theorem protects the gap. And in the second case if there are a significant number of terms in each row then we use Weyl's inequality along with a few results from random matrix theory.

\subsection{Insignificant number of noise terms}

\begin{thm}[Gershgorin circle theorem]
    Let $H=[h_{ij}]_{1\leq i, j\leq n}$ be an $n\times n$ matrix. Let $D(h_{ii}, r_i)\subset \mathbb{C}$ be the disc centered at $h_{ii}$ and having radius $r_i$, where $r_i = \sum_{j\neq i}|h_{ij}|$. Then every eigenvalue of $H$ lies in at least one of such $n$ many discs.

    Moreover, if $S_{k}$ is the union of $k$ many discs which is disjoint from the union of the other $n-k$ many discs, then $S_k$ contains exactly $k$ many eigenvalues of $H$.
\end{thm}

Now if $H$ is a diagonal matrix and $M$ is a real symmetric random matrix whose diagonal entries are zero. This $M$ can be treated as a noise matrix. Note that since $M$ is a symmetric matrix, the eigenvalues of $H+M$ are real. By the Gershgorin theorem, we may conclude that the eigenvalues of $H+M$ lie in the intervals $[h_{ii}-R_i, h_{ii} + R_i]$, where $$R_{i}=\sum_{j\neq i}|m_{ij}|.$$

Without loss of generality, let us assume that $h_{11}>h_{22}\geq h_{33}\geq \cdots\geq h_{nn}$. By the second part of the Gershgorin's theorem, if
\begin{align}\label{condition: Gershgorin}
    R_{1}+R_{i}<h_{11}-h_{ii},\;\;\;\text{for all}\;\; 1<i\leq n, 
\end{align}
then the highest eigenvalue of $H+M$ is still in the interval $[h_{11}-R_{1}, h_{11}+R_{1}]$.

However, keep in mind that $R_i$ s are random variables. Thus, to achieve a meaningful conclusion, we must analyse the probabilistic behavior of $R_i$s. This leads us to the following cases.

\textbf{Case 1:} $m_{ij}$s are bounded random variables (such as \textit{uniform} distributions) such that \eqref{condition: Gershgorin} is satisfied. We may either choose $M$ a sparse random matrix with a few big enough $m_{ij}$s such that \eqref{condition: Gershgorin} is satisfied or we may choose $M$ a non sparse random matrix with small $m_{ij}$s.

\textbf{Case 2:} $m_{ij}$s are not bounded random variables but $\mathbb{P}(R_{1}+R_{i} > h_{11}-h_{ii})=o(n)$ for all $1<i\leq n$. In that case, 
\begin{align*}
    \mathbb{P}(R_{1}+R_{ii} < h_{11} - h_{ii} \;\;\forall\;i) &= 1- \mathbb{P}(R_{1}+R_{ii} < h_{11} - h_{ii} \;\;\text{for at least}\;i)\\
    &=1-no(n) = 1-o(1).
\end{align*}

\subsection{Significant number of noise terms}

\textbf{Weyl's Inequality:} \cite{weyl1912asymptotische} Let $H$ and $M$ be two Hermitian matrices with the respective eigenvalues $\lambda_i, \mu_i$ ordered as follows; $\lambda_{1}\geq \lambda_{2}\geq \cdots\geq \lambda_{n}$ and $\mu_{1}\geq \mu_{2}\geq \cdots\geq \mu_{n}$. Let $\hat{\lambda}_{1}\geq \hat{\lambda}_{2}\geq \cdots \geq \hat{\lambda}_{n}$ be the eigenvalues of $H+M$. Then
$$\mu_{n}\leq \hat{\lambda}_{i}-\lambda_{i}\leq \mu_{1}.$$

In our case, our original Hamiltonian $H$ is a diagonal matrix and the possible noise matrix $M$ is a symmetric random matrix. If the norm of $M$ is significantly low, the spectral gap of $H$ can still be maintained even after adding $M$ to $H$. Below we discuss about adding such a random band matrix to the original Hamiltonian $H$.

\begin{definition}[Random band matrix]
    A symmetric random matrix $M_{n\times n}$ is called a random band matrix of bandwidth $b_{n}$ if 
\begin{align*}
    M_{ij}=0\;\text{for}\;|i-j|>b_{n}\;\text{and}\;|i-j|<n-b_{n}.
\end{align*}
All the other non-zero entries of $M$ are random variables with mean zero and variance $v$.
\end{definition}

\begin{thm}[\cite{khorunzhy2004spectral}, Theorem 2.1]
    Let $M$ be an $n\times n$ random symmetric band matrix of bandwidth $b_{n}$ such that $b_{n}/(\log n)^{3}\to \infty$, then the spectral norm $\|M/\sqrt{b_{n}}\|$ remains bounded with probability $1$. In particular,
    \begin{align*}
        \limsup _{n\to\infty}\|M/\sqrt{b_{n}}\|\leq 2v.
    \end{align*}
    Moreover, if $b_{n}=O(n^{\gamma})$ for some $\gamma > 0$, then $\limsup _{n\to\infty}\|M/\sqrt{b_{n}}\|= 2v.$
\end{thm}

As we see from the above theorem that asymptotically norm of a random band matrix $M$ is bounded with high probability. Thus, the spectral gap of $H$ will still be maintained after adding $M$ to $H$.

%%%%%%%%%%%%%%%%%%%%%%%%%%%%%%%%%%%%%%%%%%%%%%%%%%%%%%%%%%%%%%%%%%%%%%%%%%%%%%
\section{Quantum Computation: Exploring Braid Topologies}
\label{sec:TQC}
%%%%%%%%%%%%%%%%%%%%%%%%%%%%%%%%%%%%%%%%%%%%%%%%%%%%%%%%%%%%%%%%%%%%%%%%%%%%%%
Quantum computation relies on manipulating and entangling quantum bits, or qubits, to perform computations. Braid topologies provide a visual and intuitive framework for designing and understanding quantum algorithms. They represent entanglement patterns and braiding operations, allowing analysis and manipulation of qubit states, and are crucial for fault-tolerant quantum computation. Topological quantum computation utilizes the principles of quantum mechanics to manipulate quantum states using the system's topology. Instead of relying on local properties like traditional quantum computers, topological quantum computers encode quantum states using the system's topology. \\ The process of performing a topological quantum computation involves three steps. First, qubits are created by combining multiple anyons, such as Fibonacci anyons\cite{brennen2008should,Freedman:2006yr,Bonesteel_2005,Hormozi_2007,Kliuchnikov}, with a net overall charge or spin of zero. Second, anyons are manipulated by braiding them, allowing for performing computational operations and creating specific quantum gates. Finally, the state of the anyons is measured by fusing them together, providing information about the quantum state and enabling the extraction of computational results. \\
In the next sub-section, we will present an innovative approach to compile quantum algorithms into unique braiding patterns tailored specifically for non-Abelian quasiparticles. Specifically, we will focus on the Fibonacci anyon model in the rotated space obtained using the unitary transformation in Sec. \ref{sec:isomorphism}. To achieve this objective, we have employed a combination of the weaving method and brute force search\cite{Bonesteel_2005,Hormozi_2007}. 
\subsection{Weaving and Exhaustive Search}
In topological quantum computation, a significant challenge arises when attempting to determine whether a given braid can construct the intended unitary matrix. Conventional methods for solving this problem, which rely on brute force, exhibit exponential time complexity. However, optimization techniques such as weaving of braids have been proposed in \cite{Bonesteel_2005,Hormozi_2007} to address this issue. Weaves represent a specific subset of braids where only a single quasiparticle is involved in the movement. They have proven capable of achieving universal quantum computation, and their restricted nature makes them more technologically feasible compared to general braiding. By focusing on weaves, the problem of numerically searching for braids that approximate desired gates simplifies.

Our algorithm involves a brute force search on three-quasiparticle braids, allowing for a maximum of 56 interchanges. Through this exhaustive search, we can typically identify braids that closely approximate the desired gate, with a distance of approximately $\epsilon$, measured using the global phase invariant distance\cite{Kliuchnikov}.
The unitary operations resulting from the interweaving of three quasiparticles can be precisely described using the following form:  
\begin{eqnarray} \label{UG}
U(\{n_i\}) &=& \Tilde{\sigma}_l^{n_p}\Tilde{\sigma}_2^{n{p-1}}\ldots \Tilde{\sigma}_1^{n_2}\Tilde{\sigma}_{1}^{n_{1}}, 
\end{eqnarray}  where the subscript $l$ can be either 1 or 2, and the exponents ${n_1, n_2, ..., n_{p-1},n_{p}}$ can take values from the set ${0,\pm 1,\pm 2 \ldots \pm 5}$. The number of elementary unitary operations, also known as the braid length, is defined as $L=\sum_{i=1}^{p}|n_{i}|$.  It is important to note that to obtain the desired braid representation of an elementary gate, as shown in (\ref{UG}), we assign specific probabilities to the values $ {n_i}=\{0,\pm 1,\pm 2,\pm 3,\pm 4\}$. These probabilities are $\{0.2, 0.2, 0.5, 0.2, 0.25\}$, respectively.  To evaluate the accuracy of this approximation, we utilize the global phase invariant distance, which is defined as\cite{Kliuchnikov}:
\begin{equation}{\label{ON}}
  \epsilon= \sqrt{1 - \frac{|\Tr(U.V^{\dag})|}{2}},  
\end{equation}
here, $\Tr$ denotes the standard matrix trace, which represents the sum of the diagonal elements of a matrix. The distance $\epsilon$ provides a measure of the disparity between $U$ and $V$ while disregarding their overall phases. When a higher level of precision is required, brute force searching becomes exceedingly difficult and impractical due to its exponential complexity. Fortunately, the Solovay-Kitaev theorem \cite{Kitaev2002ClassicalAQ,nielsen2002quantum}  offers a solution. The length of the braid required for the approximation grows logarithmically with the desired accuracy.
In this article, we explore the utilization of the weaving method along with brute force techniques to manipulate braids composed of a maximum of $56$ elementary braid operations. Our objective is to find braids that closely approximate a desired target gate. Through our investigations, we consistently discovered approximate braids that achieve this goal, exhibiting a distance of approximately $10^{-3}$ as measured by the operator norm $\epsilon$ (\ref{ON}). To illustrate our findings, we present the first example, which involves searching for a quantum circuit that emulates the Hadamard gate(see in Fig. \ref{HG}):
\begin{figure}[ht]{\label{HG}}
\centering
    \begin{tikzpicture}
        \node[scale=1.0] {
            \begin{quantikz}
                \ket{0} & \gate{H}     & \meter{} & \qw
            \end{quantikz}
        };
    \end{tikzpicture}
    \caption{Hadamard Gate}
    \label{HG}
\end{figure}
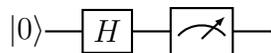

\begin{figure}[ht]
\centering
\resizebox{15cm}{2cm}{
\begin{tikzpicture}
\pic[rotate=90,
braid/.cd,
every strand/.style={ultra thick},
strand 1/.style={blue},
strand 2/.style={red},
strand 3/.style={black},
strand 4/.style={blue},
] {braid={ s_ 1^{1}s_ 1^{1}s_ 1^{1}s_ 1^{1} s_ 2 s_ 2 s_ 1 s_ 1 s_ 2^{-1} s_ 2^{-1} 
s_ 1^{-1} s_1^{-1}s_ 2^{-1} s_ 2^{-1} s_ 2^{-1} s_ 1^{-1} s_ 1^{-1} s_ 2^{-1} s_ 2^{-1} s_ 2^{-1}s_ 1  s_ 1 s_ 2  s_ 2 s_1^{-1} s_ 1^{-1} s_ 2 s_ 2 s_ 2 s_ 2 s_ 1 s_1 }};
\end{tikzpicture}}
\caption{ Braid representation of quantum Hadamard gate($H$)}\label{HGATE}
\end{figure}
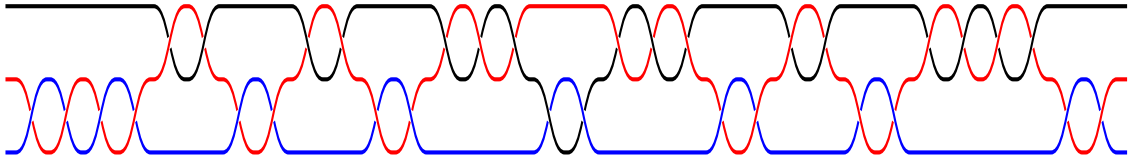

\begin{figure}[ht]
    \centering
\resizebox{15cm}{2cm}{
\begin{tikzpicture}
\pic[rotate=90,
braid/.cd,
every strand/.style={ultra thick},
strand 1/.style={blue},
strand 2/.style={red},
strand 3/.style={black},
strand 4/.style={blue},
]{braid={ s_ 1^{-1} s_1^{-1}s_ 1^{-1} s_1^{-1} s_2^{-1} s_2^{-1} s_1 s_1 s_1 s_1 s_2^{-1} s_2^{-1} s_1 s_1  s_1 s_1 s_2^{-1} s_2^{-1} s_1 s_1 s_1 s_1 s_2^{-1} s_2^{-1} s_1^{-1} s_1^{-1}s_1^{-1} s_1^{-1}}};
\end{tikzpicture}}
    \caption{The braid representation of NOT gate(with overall phase $\mathrm{i}$)}
    \label{NOT}
\end{figure}
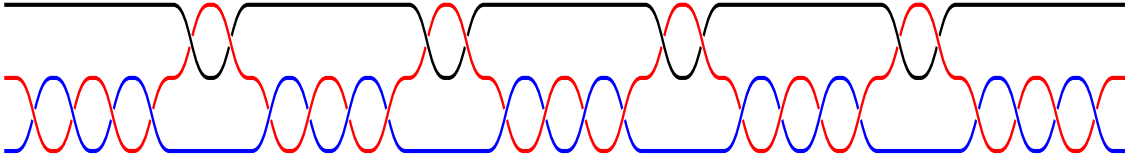

%\begin{center}
%\begin{quantikz}
%& \gate{X} &\gate{H} & \gate{X}& &&
%\end{quantikz}
%\end{center}
%\begin{tikzpicture}
%\pic[rotate=90,
%braid/.cd,
%every strand/.style={ultra thick},
%strand 1/.style={blue},
%strand 2/.style={red},
%strand 3/.style={black},
%strand 4/.style={blue},
%] {braid={ s_ 1^{-1} s_ 1^{-1} s_ 2^{-1} s_ 2^{-1} s_ 2^{-1}s_ 2^{-1} s_ 1 s_ 1 s_1 s_1 
%s_ 2^{-1} s_ 2^{-2} s_ 1 s_ 1 s_ 1 s_ 1 s_ 2  s_ 2 s_ 1 s_ 1 s_ 2 s_ 2 s_ 2 s_ 1^{-1}s_ 1^{-1} s_ 2 s_ 2  s_1^{-1} s_2 s_ 2 s_ 1^{-1} }};
%\end{tikzpicture}} %$\textit{with error of }\epsilon=6.47 \times 10^{-3}$.\\
where, the Hadamard gate, often denoted as $H$, is a fundamental quantum gate that operates on a single qubit. The Hadamard gate is represented by the following matrix:
$$H=\frac{1}{\sqrt{2}}\left(
\begin{array}{cc}
 1 & 1 \\
 1 & -1 \\
\end{array}
\right).
$$
When applied to a qubit, the Hadamard gate creates a superposition of states by transforming the basis states $\ket{0}$ and $\ket{1}$. Mathematically,
\begin{equation}
    H\ket{0}=\frac{1}{\sqrt{2}}\left(\ket{0}+ \ket{1}\right),\nonumber\\~~
        H\ket{1}=\frac{1}{\sqrt{2}}\left(\ket{0}-\ket{1}\right).\nonumber
\end{equation}
Through our investigations, we consistently discovered approximate braid is composed of $32$ elementary braiding operations and has the braidword(see in Fig. \ref{HGATE}):$$\Tilde{\sigma}_ 1^{4}  \Tilde{\sigma}_2^{2} \Tilde{\sigma}_ 1^{2} \Tilde{\sigma}_ 2^{-2} \Tilde{\sigma}_ 1^{-2} \Tilde{\sigma}_ 2^{-3}  \Tilde{\sigma}_1^{-2}  \Tilde{\sigma}_ 2^{-3}  \Tilde{\sigma}_ 1^{2} \Tilde{\sigma}_2^{2} \Tilde{\sigma}_1^{-2} \Tilde{\sigma}_2^{4} \Tilde{\sigma}_1^{2},~~ \textit{with error of}~\epsilon=6.48 \times 10^{-3}.$$\\ In the context of the provided error range, our findings surpass the results presented in \cite{Bernard-Simula}. 
Now, we present a second example involving the NOT gate.\\
$\bullet$ The NOT gate \footnote{It is named after its similarity to the classical NOT gate, which flips the value of a classical bit.}, also known as the Pauli-$X$ gate or $X$ gate, is a fundamental quantum gate that operates on a single qubit. In quantum computing, the NOT gate flips the quantum state of a qubit, effectively changing a qubit in the $\ket{0}$ to $\ket{1}$ state, and vice versa. We have found the braid representation of NOT gate i.e
$\left(
\begin{array}{cc}
 0 & 1 \\
 1 & 0 \\
\end{array}
\right)$. Here, the braid is composed of $28$ elementary braiding operations and has the braidword(shown in Fig. \ref{NOT}):
$$\Tilde{\sigma}_ 1^{-4}  \Tilde{\sigma}_2^{-2} \Tilde{\sigma}_ 1^{4}  \Tilde{\sigma}_ 2^{-2} \Tilde{\sigma}_ 1^4 \Tilde{\sigma}_ 2^{-2}  \Tilde{\sigma}_1^{4}  \Tilde{\sigma}_ 2^{-2} \Tilde{\sigma}_ 1^{-4}, \textit{with error of }~\epsilon=2.19\times 10^{-3}.$$
Furthermore, there are two significant single-qubit gates: the $T$ gate and the $S$ gate. These gates introduce specific phase shifts to qubit states.

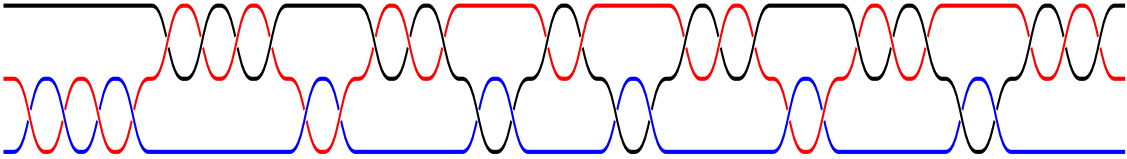
\begin{figure}[ht]
    \centering
\resizebox{15cm}{2cm}{
\begin{tikzpicture}
\pic[rotate=90,
braid/.cd,
every strand/.style={ultra thick},
strand 1/.style={blue},
strand 2/.style={red},
strand 3/.style={black},
strand 4/.style={blue},
] {braid={ s_ 1^{-1} s_ 1^{-1} s_ 1^{-1} s_ 1^{-1} s_ 2 s_ 2 s_ 2 s_ 2 s_ 1 s_1 
s_ 2^{-1} s_ 2^{-1} s_ 2^{-1} s_ 1^{-1} s_ 1^{-1} s_ 2 s_ 2  s_ 1^{-1}s_ 1^{-1} s_ 2 s_ 2 s_ 2 s_1 s_1 s_ 2^{-1}s_ 2^{-1}s_ 2^{-1} s_ 1^{-1} s_1^{-1} s_ 2 s_ 2 s_ 2}};
\end{tikzpicture}}
    \caption{The braid representation of $S$ gate }\label{SGATE}
\end{figure}

\begin{figure}[ht]
    \centering
\resizebox{15cm}{2cm}{
\begin{tikzpicture}
\pic[rotate=90,
braid/.cd,
every strand/.style={ultra thick},
strand 1/.style={blue},
strand 2/.style={red},
strand 3/.style={black},
strand 4/.style={blue},
] {braid={ s_ 1 s_ 1 s_ 1 s_ 2 s_2 s_ 1^{-1} s_ 1^{-1} s_ 2^{-1} s_2^{-1} 
s_ 2^{-1} s_ 2^{-1} s_ 1  s_ 2 s_ 2 s_ 2 s_ 1  s_ 1 s_ 2^{-1} s_ 2^{-1} s_ 1 s_ 1 s_2^{-1} s_2^{-1} s_ 1^{-1} s_ 1^{-1} s_ 1^{-1} s_ 1^{-1} s_ 2^{-1}s_ 2^{-1} s_ 2^{-1} s_2^{-1} s_ 1^{-1} s_1^{-1} s_ 2^{-1}s_ 2^{-1}}};
\end{tikzpicture}}
    \caption{The braid representation of quantum $T$ gate}
    \label{TGATE}
\end{figure}
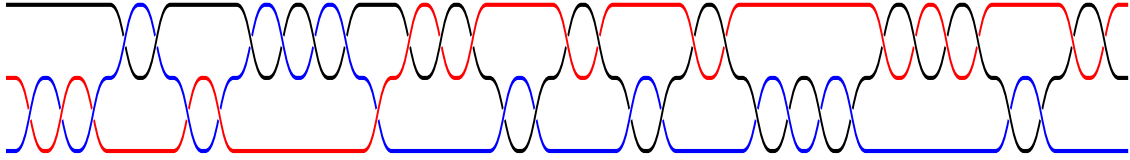

\begin{figure}[ht]
  \centering
  \subfigure(a){\begin{tikzpicture}
        \node[scale=1.1] {
            \begin{quantikz}
                \ket{0} & \gate{H}  & \gate{i X}  & \gate{H} & \meter{} & \qw
            \end{quantikz}
        };
    \end{tikzpicture}} 
  \subfigure(b){\resizebox{8cm}{5cm}{
\begin{tikzpicture}
\draw[rotate=270,blue, very thick] (-2.,-4) -- (-2.,-4.)
        arc [x radius = 7.3em, y radius = 7.8em, start angle = 180, end angle = 0];
        \draw[rotate=270,black, very thick] (-1,-4.) -- (-1,-4.)
        arc [x radius = 4.9em, y radius = 6.2em, start angle = 180, end angle = 0];
        \draw[rotate=270,red, very thick] (0,-4) -- (0,-4)
        arc [x radius = 2.4em, y radius = 3.2em, start angle = 180, end angle = 0];
        \pic[rotate=90, braid/.cd,
every strand/.style={ultra thick},
strand 1/.style={red},
strand 2/.style={black},
strand 3/.style={blue},
] at (-36.5,0) {braid={s_ 1^{1}s_ 1^{1}s_ 1^{1}s_ 1^{1} s_ 2 s_ 2 s_ 1 s_ 1 s_ 2^{-1} s_ 2^{-1} 
s_ 1^{-1} s_1^{-1}s_ 2^{-1} s_ 2^{-1} s_ 2^{-1} s_ 1^{-1} s_ 1^{-1} s_ 2^{-1} s_ 2^{-1} s_ 2^{-1}s_ 1  s_ 1 s_ 2  s_ 2 s_1^{-1} s_ 1^{-1} s_ 2 s_ 2 s_ 2 s_ 2 s_ 1 s_1}};
 \pic[rotate=270,
braid/.cd,
every strand/.style={ultra thick},
strand 1/.style={red},
strand 2/.style={black},
strand 3/.style={blue},
]at (-3.6,-1.98) {braid={s_ 1^{-1} s_1^{-1}s_ 1^{-1} s_1^{-1} s_2^{-1} s_2^{-1} s_1 s_1 s_1 s_1 s_2^{-1} s_2^{-1} s_1 s_1  s_1 s_1 s_2^{-1} s_2^{-1} s_1 s_1 s_1 s_1 s_2^{-1} s_2^{-1} s_1^{-1} s_1^{-1}s_1^{-1} s_1^{-1}}};
\draw[rotate=90,blue, very thick] (-7.,32) -- (-7.,32)
        arc [x radius = 3.6em, y radius = 3.5em, start angle = 180, end angle = 0];
        \draw[rotate=90,black, very thick] (-8,32) -- (-8,32)
        arc [x radius = 6.1em, y radius = 6em, start angle = 180, end angle = 0];
        \draw[rotate=90,red, very thick] (-9,32) -- (-9,32)
        arc [x radius = 8.5em, y radius = 8.1em, start angle = 180, end angle = 0];
  \pic[rotate=90,
braid/.cd,
every strand/.style={ultra thick},
strand 1/.style={red},
strand 2/.style={black},
strand 3/.style={blue},
] at (-32,-9) {braid={s_ 1^{1}s_ 1^{1}s_ 1^{1}s_ 1^{1} s_ 2 s_ 2 s_ 1 s_ 1 s_ 2^{-1} s_ 2^{-1} 
s_ 1^{-1} s_1^{-1}s_ 2^{-1} s_ 2^{-1} s_ 2^{-1} s_ 1^{-1} s_ 1^{-1} s_ 2^{-1} s_ 2^{-1} s_ 2^{-1}s_ 1  s_ 1 s_ 2  s_ 2 s_1^{-1} s_ 1^{-1} s_ 2 s_ 2 s_ 2 s_ 2 s_ 1 s_1}};
 \end{tikzpicture}}} 
\caption{(a) Quantum circuit diagram of single qubit gates (b)The braid representation of single qubit gates}\label{SQG}
\end{figure}
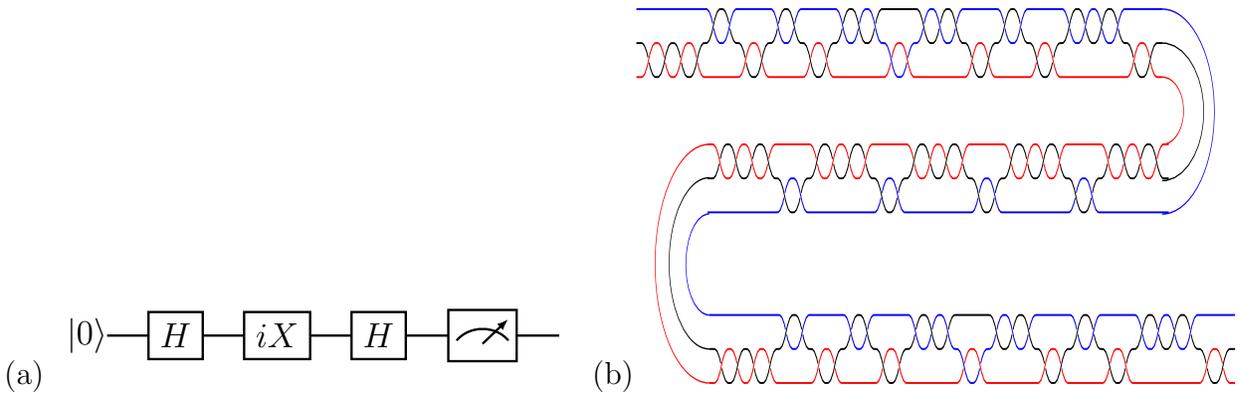

$\bullet$ $S$ gate ($\frac{\pi}{2}$ phase shift): The $S$ gate, also known as the $\frac{\pi}{2}$ gate or the phase gate, is a quantum gate that applies a phase shift of $\frac{\pi}{2}$ radians ($90$ degrees) to a qubit state. It is represented by the following matrix:
$$S=\left(
\begin{array}{cc}
 1 & 0 \\
 0 & \mathrm{i} \\
\end{array}
\right)
.$$ The braid consists of a total of 32 elementary braiding operations and is represented by the braidword in Fig. \ref{SGATE}  $$\Tilde{\sigma}_ 1^{-4} \Tilde{\sigma}_2^{4} \Tilde{\sigma}_ 1^{2} \Tilde{\sigma}_ 2^{-3} \Tilde{\sigma}_ 1^{-2} \Tilde{\sigma}_ 2^{2}  \Tilde{\sigma}_1^{-2}  \Tilde{\sigma}_ 2^{3}  \Tilde{\sigma}_ 1^{2} \Tilde{\sigma}_2^{-3} \Tilde{\sigma}_1^{-2} \Tilde{\sigma}_2^{3}.$$ The estimated error for this braid is $\epsilon = 3.9\times 10^{-3}$. It should be noted that the braid length associated with the $S$ gates is smaller compared to the findings presented in \cite{Bernard-Simula}.\\
$\bullet$ $T$ gate ($\frac{\pi}{4}$ phase shift): The $T$ gate, also known as the $\frac{\pi}{4}$ gate, is a quantum gate that applies a phase shift of $\frac{\pi}{4}$ radians ($45$ degrees) to a qubit state. It is represented by the following matrix:
$$T=\left(
\begin{array}{cc}
 1 & 0\\
 0 & \exp{(\frac{\mathrm{i} \pi}{4} )} \\
\end{array}
\right).$$
When applied to a qubit, the $T$ gate leaves the \ket{0} state unchanged, while introducing a phase shift of $\frac{\pi}{4}$ to the \ket{1} state. The braid is constructed using a sequence of 31 elementary braiding operations, resulting in the braidword (See Fig. \ref{TGATE}):
$$\Tilde{\sigma}_ 1^{3}  \Tilde{\sigma}_2^{2} \Tilde{\sigma}_ 1^{-2}  \Tilde{\sigma}_ 2^{-4} \Tilde{\sigma}_ 1 \Tilde{\sigma}_ 2^{3}  \Tilde{\sigma}_1^{2}  \Tilde{\sigma}_ 2^{-2}  \sigma_ 1^{2}\Tilde{\sigma}_ 2^{-2}  \Tilde{\sigma}_1^{-4}  \Tilde{\sigma}_ 2^{-4}  \Tilde{\sigma}_ 1^{-2}\Tilde{\sigma}_ 2^{-2}, \textit{with error of }\epsilon=9.63\times 10^{-3}.$$\\
In the realm of topological quantum computation, we have made progress in achieving single-qubit gates(see in Fig.\ref{SQG}). However, when it comes to two-qubit gates, there are various techniques that have been discussed in detail, such as effective braiding weaves, injection weaves, and single-quasiparticle weaves \cite{Bonesteel_2005, Hormozi_2007} for efficient gate compilation. To elaborate, effective braiding weaves are utilized to approximate the process of interchanging quasiparticles in a braided system. 
These weaves are applicable only for even values of a parameter denoted as ${n_{i}}$ in (\ref{UG}). The construction of these weaves involves an exhaustive search to find suitable ones, followed by applying the Solovay-Kitaev algorithm \cite{Kitaev2002ClassicalAQ, nielsen2002quantum} to enhance their accuracy. On the other hand, injection weaves are employed to permute quasiparticles while keeping their underlying quantum properties intact. These weaves approximate the identity operation and require the weft particle to end up at the bottom of the weave. It is worth noting that injection weaves can efficiently compile arbitrary controlled rotations of the target qubit, including a controlled-NOT gate. While initially discussed in \cite{Bonesteel_2005}, it becomes evident that these principles can be extended to encompass two-qubit gates as well.

%%%%%%%%%%%%%%%%%%%%%%%%%%%%%%%%%%%%%%%%%%%%%%%%%%%%%%%%%%%%%%%%%%%%%%%%%%%%%%
\section{Outlook}
\label{sec:outlook}
\begin{figure}[h]
     \centering
     \includegraphics[width=8cm]{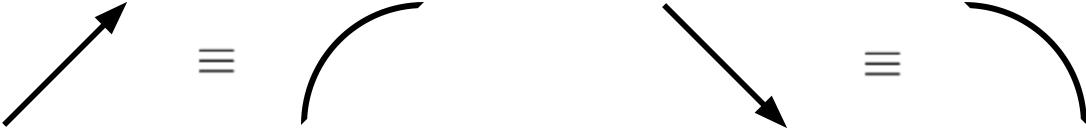}
     \caption{Mapping to go between the arrow representation of the Dyck paths to the loop representation.}
      \label{fig:arrow2loop}
 \end{figure}

\begin{figure}[h]
     \centering
     \includegraphics[width=8cm]{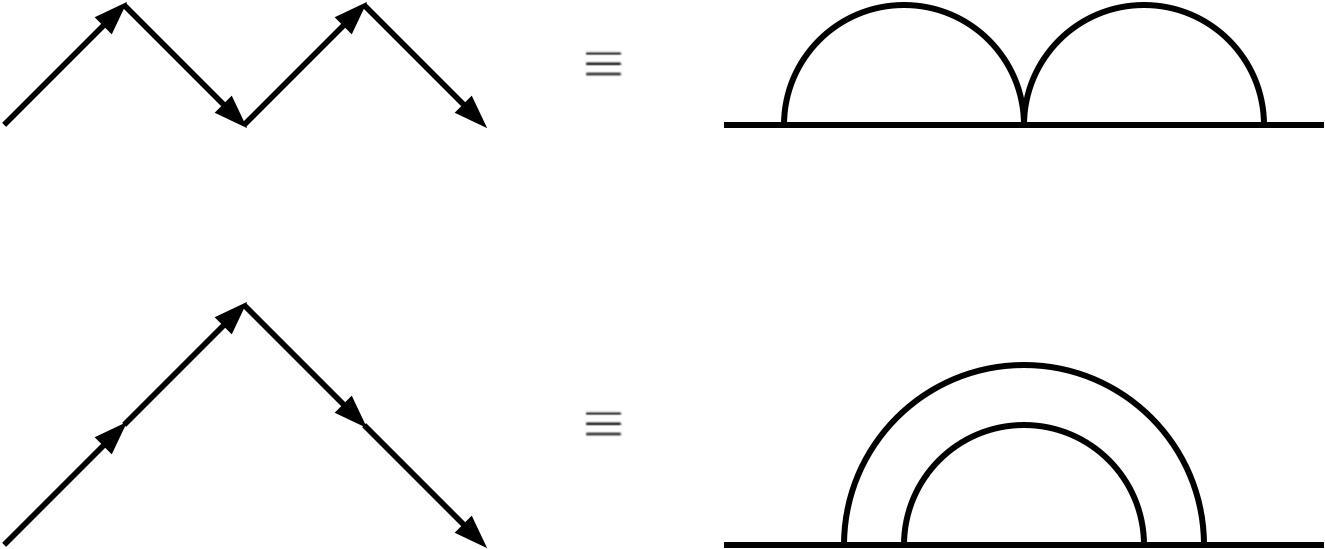}
     \caption{The length 4 Dyck paths in the loop representation.}
      \label{fig:loopstatesN4}
 \end{figure}

\begin{figure}[h]
     \centering
     \includegraphics[width=8cm]{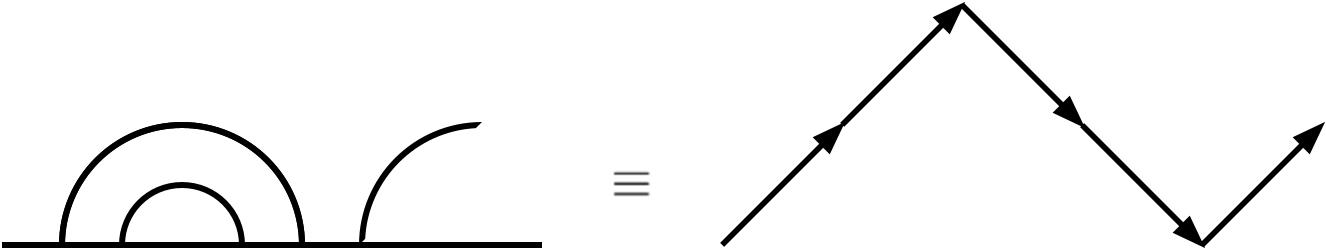}
     \caption{A length 5 link state in the loop and arrow representation.}
      \label{fig:linkstateN5}
 \end{figure}

%%%%%%%%%%%%%%%%%%%%%%%%%%%%%%%%%%%%%%%%%%%%%%%%%%%%%%%%%%%%%%%%%%%%%%%%%%%%%%
The non-Abelian anyons from the IRR's of the $SU(2)_k$ theories provide a systematic way to construct non-local and unitary braid group representations via the Jones form. Thus the fusion spaces corresponding to each of these anyons carries a representation of the Temperley-Lieb algebra as well. It is then natural to probe the relation between them and spin chain states such as the Dyck paths studied in this work. Indeed such a correspondence is expected {\it via} the generalisation of the Dyck paths known as {\it link states} \cite{Ridout2012StandardMI}. To understand the latter we take a different look at the Dyck paths in terms of half-loops on the line \cite{Gier2002LoopsMA,ZinnJustin2009SixvertexLA, Francesco1996MeandersAT} which lead to the {\it loop representation} of the Temperley-Lieb algebra and consequently the braid group. To obtain the loop representations the `up' and `down' steps are identified with `arcs' as shown in Fig. \ref{fig:arrow2loop}.
%%%%%%%%%%%%%%%%%%%%%%%%%%%%%%%%%%%%%%%%%%%%%%%%%%%%%%%
%%%%%%%%%%%%%%%%%%%%%%%%%%%%%%%%%%%%%%%%%%%%%%%%%%%%%%%
Using this replacement we can write down Dyck paths as loops (See Fig. \ref{fig:loopstatesN4} for the loop representation of the length 4 Dyck paths).
%%%%%%%%%%%%%%%%%%%%%%%%%%%%%%%%%%%%%%%%%%%%%%%%%%%%%%%
%%%%%%%%%%%%%%%%%%%%%%%%%%%%%%%%%%%%%%%%%%%%%%%%%%%%%%%
Link states are those states where some of the half-loops in the loop representation are incomplete or equivalently in the arrow representation they are those configurations where the number of `up' arrows are not the same as the number of `down' arrows. These states can now be defined on chains with an odd number of steps as well. For example a length 5 link state is shown in Fig. \ref{fig:linkstateN5}.
%%%%%%%%%%%%%%%%%%%%%%%%%%%%%%%%%%%%%%%%%%%%%%%%%%%%%%%
%%%%%%%%%%%%%%%%%%%%%%%%%%%%%%%%%%%%%%%%%%%%%%%%%%%%%%%
The Temperley-Lieb algebra can be represented on such link states \cite{Ridout2012StandardMI,Westbury1995} and by counting the dimensions of these representations we can arrive at correspondences (unitary transformation) to the fusion basis of other $SU(2)_k$ anyonic systems. This will then act as the guide to construct the corresponding local spin chains for these systems. Following this we can also check for the universal computation properties of these systems and study if the unitary rotation provides any distinct advantage as it did for the Fibonacci case. We intend to address these problems in a future work.

Another avenue to explore is an intriguing relationship between Chern-Simons theory and two-dimensional conformal field theory\cite{Witten:1988hf}. This relationship suggests that the Hilbert space of states in $SU(2)$ Chern-Simons theory, with a coupling constant $k$, on a three-manifold $S^2 \times R$, can be fully characterized by the conformal blocks of the $SU(2)_k$ Wess-Zumino conformal theory on a two-sphere ($S^2$) with punctures. Each puncture is associated with a spin value $j$ ranging from $\frac{1}{2}$ to $\frac{k}{2}$.  
By considering the fusion of $2N$ primary fields, each possessing a specific spin value ($j=\frac{1}{2}$) on the external lines. It becomes evident that, for sufficiently large values of the level $k$, the dimension of the Hilbert space (i.e., the total number of conformal blocks) of a two-sphere hosting $2N$ punctures, the Verlinde formula reads \cite{Verlinde:1988sn,Witten:1988hf}:
\begin{equation}
\lim_{k\rightarrow \infty}dim(\mathcal{H}_{S^2})=\lim_{k\rightarrow \infty}\frac{2}{k+2}\sum_{r=0}^{k}\frac{\sin{\left(\frac{2(r+1)\pi}{k+2}\right)^{2N}}}
{\sin{\left(\frac{(r+1)\pi}{k+2}\right)^{2N-2}}}=C_{N},
\end{equation}
where $C_{N}$ represent the Catalan numbers (Sec. \ref{sec:isomorphism}). Remarkably, the dimension of the Temperley-Lieb algebra $TL_N$ corresponds to the Catalan numbers, establishing a direct connection to the dimension of the Hilbert space in this particular scenario. This becomes especially interesting in the context of recent works \cite{Bellette2018FusionAM,Gainutdinov2016FusionAB} using which we can think of obtaining a functor between the Temperley-Lieb category and the fusion category of anyons. This would give a more formal footing to the correspondences studied in this work and will help generalise our results to other anyon systems. 

The Motzkin spin chain \cite{Bravyi2012CriticalityWF,Movassagh2014PowerLV,Sugino2017AreaLV,Sugino2019HighlyEQ} is another system exhibiting exotic entanglement entropy properties like the Fredkin model. It would be interesting to check if the Motzkin walks also support topological quantum computation in the same manner as the Dyck walks governed by the Fredkin moves.

%%%%%%%%%%%%%%%%%%%%%%%%%%%%
\section*{Acknowledgements}
%%%%%%%%%%%%%%%%%%%%%%%%%%%%
IJ is partially supported by DST's INSPIRE Faculty Fellowship through the grant \\ DST/INSPIRE/04/2019/000015. The work of VKS is supported by “Tamkeen under the NYU Abu
Dhabi Research Institute grant CG008 and ASPIRE Abu Dhabi under Project AARE20-
336”.

%%%%%%%%%%%%%%%%%%%%%%%%%%%%%%
\appendix
%%%%%%%%%%%%%%%%%%%%%%%%%%%%%%%
%%%%%%%%%%%%%%%%%%%%%%%%%%%%%%%%%%%%%%%%%%%%%%%%%%%%%%%%%%
\section{$\mathcal{B}_N$ representations on Dyck paths}
\label{app:BNDyckPaths}
%%%%%%%%%%%%%%%%%%%%%%%%%%%%%%%%%%%%%%%%%%%%%%%%%%%%%%%%%%
Though we do not use the Dyck path representations of the braid group for arbitrary $N$ in the main text, we include them here to make the article self-contained. The expressions that follow are adapted from \cite{Jana2022TopologicalQC} where they are worked out in more detail. As mentioned in the main text, the representations for arbitrary $N$ are built out of the IRR's $\mathcal{B}_2$ and $\mathcal{B}_3$. These are apparent from their expressions,
\begin{eqnarray}
    e_1~\ket{1_1\cdots} & = & x~\ket{1_1\cdots}, \nonumber \\
    e_1~\ket{\tau_1\cdots} & = & 0, \nonumber \\
    e_2~\ket{1_1\tau_2\cdots} & = & \frac{x^2-1}{x}~\ket{1_1\tau_2\cdots} + \frac{\sqrt{x^2-1}}{x}~\ket{\tau_1\tau_2\cdots}, \nonumber \\
    e_2~\ket{\tau_1\tau_2\cdots} & = & \frac{\sqrt{x^2-1}}{x}~\ket{1_1\tau_2\cdots} + \frac{x^2-1}{x}~\ket{\tau_1\tau_2\cdots}, \nonumber \\
    e_2~\ket{\tau_11_2\cdots} & = & 0,
\end{eqnarray}
for the left generators,
\begin{eqnarray}
    e_{N-1}~\ket{\cdots 1_{N-3}\tau_{N-2}} & = & 0, \nonumber \\
    e_{N-1}~\ket{\cdots \tau_{N-3}1_{N-2}} & = & \frac{x^2-1}{x}~\ket{\cdots \tau_{N-3}1_{N-2}} + \frac{\sqrt{x^2-1}}{x}~\ket{\cdots \tau_{N-3}\tau_{N-2}}, \nonumber \\
    e_{N-1}~\ket{\cdots \tau_{N-3}\tau_{N-2}} & = & \frac{\sqrt{x^2-1}}{x}~\ket{\cdots \tau_{N-3}1_{N-2}} + \frac{x^2-1}{x}~\ket{\cdots \tau_{N-3}\tau_{N-2}},
\end{eqnarray}
for the right generator and,
\begin{eqnarray}
    e_i~\ket{\cdots 1_{i-2}\tau_{i-1}1_i\cdots} & = & x~\ket{\cdots 1_{i-2}\tau_{i-1}1_i\cdots}, \nonumber \\
    e_i~\ket{\cdots 1_{i-2}\tau_{i-1}\tau_i\cdots} & = & 0, \nonumber \\
    e_i~\ket{\cdots \tau_{i-2}\tau_{i-1}1_i\cdots} & = & 0, \nonumber \\
    e_i~ \ket{\cdots \tau_{i-2}1_{i-1}\tau_i\cdots} & = & \frac{x^2-1}{x}~\ket{\cdots \tau_{i-2}1_{i-1}\tau_i\cdots} + \frac{\sqrt{x^2-1}}{x}~\ket{\cdots \tau_{i-2}\tau_{i-1}\tau_i\cdots}, \nonumber \\
    e_i~ \ket{\cdots \tau_{i-2}\tau_{i-1}\tau_i\cdots} & = & \frac{\sqrt{x^2-1}}{x}~\ket{\cdots \tau_{i-2}1_{i-1}\tau_i\cdots} + \frac{x^2-1}{x}~\ket{\cdots \tau_{i-2}\tau_{i-1}\tau_i\cdots}, \nonumber \\
\end{eqnarray}
for the bulk generators. Finally to obtain the expressions on the Dyck paths we use the correspondence between the Fibonacci fusion basis and the length 4 Dyck paths in \eqref{eq:Fibstates2Dyckpaths}. The generators can then be written in terms of Pauli $X$ and $Z$ operators using,
\begin{equation}
    \ket{u}\bra{u} = \frac{1+Z}{2},~\ket{d}\bra{d} = \frac{1-Z}{2},~\ket{u}\bra{d} = \frac{X+\mathrm{i}Y}{2},~\ket{d}\bra{u} = \frac{X-\mathrm{i}Y}{2}.
\end{equation}

\bibliographystyle{unsrt}
\normalem
\bibliography{refs}

\end{document}